\documentclass{article}

\usepackage{PRIMEarxiv}

\usepackage[utf8]{inputenc} 
\usepackage[T1]{fontenc}    
\usepackage{hyperref}       
\usepackage{url}            
\usepackage{booktabs}       
\usepackage{amsfonts}       
\usepackage{nicefrac}       
\usepackage{microtype}      
\usepackage{lipsum}
\usepackage{fancyhdr}       
\usepackage{graphicx}       
\usepackage{bm}
\usepackage{subfig}
\usepackage{float}
\usepackage{amsmath}
\usepackage{multirow}
\graphicspath{{media/}}     

\title{Physics-informed neural networks of the Saint-Venant equations for downscaling a large-scale river model
}

\author{
 Dongyu Feng \\
  Pacific Northwest National Laboratory\\
  Richland, WA 99354 \\
  \texttt{dongyu.feng@pnnl.gov} \\
   \And
 Zeli Tan \\
  Pacific Northwest National Laboratory\\
  Richland, WA 99354 \\
  \texttt{zeli.tan@pnnl.gov} \\
  \And
 QiZhi He \\
  University of Minnesota\\
  Minneapolis, MN 55455 \\
  \texttt{qzhe@umn.edu} \\
}

\begin{document}
\maketitle

\begin{abstract}
Large-scale river models are being refined over coastal regions to improve the scientific understanding of coastal processes, hazards and responses to climate change. However, coarse mesh resolutions and approximations in physical representations of tidal rivers limit the performance of such models at resolving the complex flow dynamics especially near the river-ocean interface, resulting in inaccurate simulations of flood inundation. In this research, we propose a machine learning (ML) framework based on the state-of-the-art physics-informed neural network (PINN) to simulate the downscaled flow at the subgrid scale. First, we demonstrate that PINN is able to assimilate observations of various types and solve the one-dimensional (1-D) Saint-Venant equations (SVE) directly. We perform the flow simulations over a floodplain and along an open channel in several synthetic case studies. The PINN performance is evaluated against analytical solutions and numerical models. Our results indicate that the PINN solutions of water depth have satisfactory accuracy with limited observations assimilated. In the case of flood wave propagation induced by storm surge and tide, a new neural network architecture is proposed based on Fourier feature embeddings that seamlessly encodes the periodic tidal boundary condition in the PINN's formulation. 
Furthermore, we show that the PINN-based downscaling can produce more reasonable subgrid solutions of the along-channel water depth by assimilating observational data. The PINN solution outperforms the simple linear interpolation in resolving the topography and dynamic flow regimes at the subgrid scale. This study provides a promising path towards improving emulation capabilities in large-scale models to characterize fine-scale coastal processes. 
\end{abstract}


\section{Introduction}
\label{s:intro}

The population growth near coastal regions increases human exposure to natural hazards of hurricanes and flooding \cite{tellman2021satellite}. 
Coastal inundation area changes greatly during extreme flooding events. The flow and fluxes at the terrestrial-aquatic interface or tidal rivers are governed by river discharge, tide and storm surge, as well as their complex physical interactions \cite{hoitink2016tidal}, which could also impact sediment dynamics and biogeochemical processes \cite{lamb2012backwater,ward2020representing}. For example, during landfalling with tropical cyclones bringing heavy precipitation, the flood wave propagation from downstream storm surge and tide impedes the upstream discharge and this backwater effect in turn modulates the water level \cite{dykstra2020propagation,gori2020assessing,dykstra2021role}.
Such flood event is referred as compound flooding as multiple mechanisms can occur simultaneously or in close successions \cite{moftakhari2017compounding,eilander2020effect}. 
Climate warming will further exacerbate risks of such hazards because it will create more dynamic flooding zones in river systems over the low-lying regions \cite{santiago2019comprehensive} by intensifying extreme storm surge and precipitation \cite{alfieri2016increasing,marsooli2019climate}, accelerating sea level rise (SLR) \cite{kulp2019new} and enhancing tidal dynamics \cite{talke2020changing}. 
It thus becomes crucial to characterize the complex interactions between these processes in a warming climate. 

The mitigation of compound flood risks requires the understanding of tidal river dynamics, which are affected by the nonlinear interaction of hydrological and hydrodynamical processes, river geometry and local basin characteristics \cite{haigh2020tides}. Beyond the traditional harmonic tidal analysis, there is a growing application of using numerical models to study the compounding dynamics. For example, hydraulic or hydrodynamic models are configured in tidal estuaries to simulate coastal inundation and flood wave propagation in river networks at the event scale \cite{zhang2020simulating,helaire2020present,xiao2021characterizing} or under a changing climate \cite{barnard2019dynamic}. The upstream and coastal boundary conditions of these models are provided by hydrological and coastal models, respectively. There are also studies that apply more efficient river hydrological models \cite{ikeuchi2017compound}, coupled with global tide and surge models \cite{muis2016global,muis2020high}, in larger-scale risk assessments \cite{eilander2020effect}. However, these modeling studies usually have not fully utilized gauged measurements that are abundant in tidal rivers and high-resolution remote sensing data. Such data, if appropriately assimilated, will likely improve the model performance, especially for the large-scale models that under-resolve the river dynamics.

Large-scale river models are a major component of Earth System Models (ESMs) that are the primary tool for climate change research. Such models have been used extensively in global water cycle simulations \cite{leung2020introduction} and large-scale flood risk assessments \cite{towner2019assessing}. Although local river models are frequently used to assess the compound flooding induced by pluvial, fluvial and coastal processes \cite{bakhtyar2020new}, the large-scale models are preferred at regional to global scales \cite{ikeuchi2017compound,yamazaki2011physically}.
However, the approximations in the governing physics and the deficiency in mesh resolutions limit these models in representing the local flooding processes, particularly near the highly dynamic terrestrial-aquatic interface. 
While the large-scale river models can conserve water mass and simulate reasonable river discharge at the global scale \cite{li2013physically}, these reduced-physics models employ simplified forms for momentum propagation, which requires less data in river channel topography and lower computational cost than dynamic models \cite{hodges2013challenges}. These approximations inevitably limit the representation of physical characteristics of dynamic systems \cite{santiago2019comprehensive}. 

Specifically, in a one-dimensional (1-D) river system, the flow dynamics is governed by the 1-D Saint-Venant equations (SVE). Solving SVE in dynamic river models is computationally demanding, because explicit numerical schemes usually require shorter time steps to converge and implicit schemes need to evaluate SVE equations multiple iterations at every time step \cite{cunge1980practical,liu2014applying,tompson2017accelerating}. While it appears feasible to apply SVE to a large scale at high spatial resolutions with the advancement of high performance computing (HPC), such dynamic models require high-quality river geometry fields. This requirement cannot be met by the large-scale river models \cite{wood2011hyperresolution}. 
As a result, approximations must be made to SVE to relax model requirements. 
For example, for the two common SVE approximations, the diffusive wave equation neglects the Eulerian inertial acceleration and assumes that gravity, friction and pressure dominate the momentum, while the local inertial equation \cite{bates2010simple} only neglects the advection term.  
These assumptions, however, are likely not valid for flood-prone river systems. In a flat sloping or tidally-influenced river system, flood wave dominates momentum propagation as velocity changes rapidly with space and time. The inertial approximation reduces the flood propagation speed, attenuates the water surface gradient \cite{de2013applicability} and underestimates the physical characteristics such as the raising and recession of limbs \cite{shih2018studying}. 

Even with simplified physics, large-scale river models have to apply low-resolution (5 km $\sim$ 25 km) meshes to accommodate the high computational cost in large-scale applications. For example, the river component of Energy Exascale Earth System Model (E3SM) applies a resolution of 12.5 km \cite{caldwell2019doe}, which is too coarse to resolve the flow dynamics induced by the change of flow regimes over various topographical features. Additionally, in such models, flood inundation over a floodplain is usually simulated using mass-conserved schemes that assume no between-cell exchange of inundated water \cite{luo2017modeling,yamazaki2012analysis}. Such schemes, despite achieving satisfied estimation of flood fraction within a cell, cannot represent within-cell heterogeneity of water depth, because the latter would never be known without resolving subgrid flow dynamics. 

While extensive efforts have been made to develop efficient numerical schemes to represent a more dynamic river system in large-scale models \cite{yu2017consistent,yu2020new}, the alternative is to derive downscaled solutions at the subgrid scale in regions of interest using statistical or machine learning (ML) approaches.
Downscaling coarse-scale numerical models or forcing data to a fine-scale solution is an established research area in climate and ocean modeling \cite{burger1996expanded,maraun2018statistical,trotta2021relocatable}. The downscaling is often classified into two categories: statistical downscaling and dynamic downscaling. The statistical downscaling usually first builds a statistical relationship between the coarse-scale model and local data and then uses the relationship to produce high spatial- or temporal-resolution outputs  \cite{wilby2013statistical}. 
In the dynamic downscaling, a local high-resolution model simulation is performed with inputs extrapolated from large-scale processes (e.g., boundary conditions from a large-scale model) to produce high-resolution outputs directly \cite{xue2014review}. The application of downscaling to river modeling is still limited. Most previous studies only focused on downscaling a single or grouped gauged measurements to a higher temporal resolution statistically for assessing the local climate change impacts \cite{tisseuil2010statistical,moghim2022downscaling}, or downscaling two-dimensional inundation maps by overlaying water depth computed from a coarse-resolution model or satellite data onto a high-resolution digital elevation model (DEM) \cite{fluet2015development,bermudez2020robust}. To our knowledge, the only method to dynamically downscale the 1-D channel flow is to linearly interpolate the modeled water depth at each grid node and available gauged observations across in-between cross-sections \cite{schumann2014downscaling}. This method does not require channel topology information and cannot resolve the spatially-varied flow regimes within a grid cell. To fill this gap, this work develops a ML-based downscaling method that efficiently assimilates available observations and computes subgrid solutions.     

Over recent years, the availability of water depth data increases rapidly as such data are consistently collected from in-situ measurements at gauging stations as well as by remote sensing \cite{elkhrachy2022flash}, such as satellite radar altimetry \cite{durand2016intercomparison}. 
In-situ data are continuous in time and sparse in space \cite{nielsen2007river}. In contrast, remote sensing data, while measured at large time intervals, can cover great spatial domains at high resolutions \cite{schumann2014downscaling}.  
For example, the upcoming Surface Water and Ocean Topography (SWOT) will enable direct measurements of water depth at a spatial resolution of 50 m for all rivers wider than 100 m \cite{biancamaria2016swot,domeneghetti2018characterizing}. Such data could potentially be directly incorporated into ML models to obtain downscaled solutions \cite{zhu2021downscaling}. Remote sensing and gauged observation data, along with numerical solutions, provide the basis for the development of ML-based methods.

The ML technique has been explored to alleviate the limitations in the river modeling from various aspects. For example,
ML models are employed as surrogates or emulators of hydrological models to perform rapid simulations for real-time predictions and model calibrations \cite{zhang2020hydrological}. ML models are also proposed to predict hydrograph \cite{zhang2020hydrological}, maximum water levels \cite{berkhahn2019ensemble}, infiltration \cite{crompton2019emulation}, as well as real-time inundation maps \cite{qian2019physics,guo2021data}. 
Despite differences in implementation, these methods, such as regression trees, random forests and neural networks, all treat the ML model as a black box. Model predictions are made based on a set of input variables, such as precipitation and streamflow. Additionally, training an emulator is usually expensive as it requires abundant including observations or multiple realizations of high-resolution numerical simulations. 
As a result, some alternative ML algorithms are proposed to replace computationally demanding components of a numerical scheme with a computationally efficient ML model. For example, the convolutional neural network (CNN) is used to solve the Euler equation for iterative pressure correction in the velocity update algorithm \cite{tompson2017accelerating}. Similarly, artificial neural networks are used to estimate the velocity in the inertial wave equation \cite{jamali2021machine}. However, these ML models are still computationally demanding for training and also they require accurate training data to avoid bad numerical solutions.

Emerged recently are the state-of-the-art physics-informed ML models. Such models are trained by enforcing the relevant physics. Data, mathematical models and domain knowledge are seamlessly integrated and implemented through regression networks \cite{karniadakis2021physics}.
Physics-informed neural networks (PINN) solve partial differential equations (PDEs) using feed-forward neural network architectures and respect the corresponding physical laws \cite{raissi2019physics}. 
The unknown solutions of governing equations are inferred from the initial and boundary data of the state variables.  
The ML models trained with physical constraints require much less data than traditional ML emulators. This approach thus greatly alleviates the limitations of ML in model training caused by data scarcity.  
The PINN algorithm can also be used in inverse problems to estimate parameters of a PDE with observations of state variables provided \cite{kissas2020machine,he2021physics}, as well as in developing surrogate models from sparse data \cite{karpatne2017physics,sun2020surrogate}. 
Specialized network architectures have been developed for PINN to meet specific physical laws \cite{leung2022nh}, accelerate the training process \cite{yu2022gradient} and improve the model accuracy \cite{de2021hyperpinn}. As the result, many new variants of PINN were introduced recently, such as multiphysics PINN \cite{he2020physics}, Parareal PINN \cite{meng2020ppinn,shukla2021parallel}, and DeepONet \cite{lu2021learning}. For a comprehensive review of PINN, please refer to \cite{karniadakis2021physics}. 

Due to the illustrated strengths, PINN has been explored extensively in solving forward, inverse and data assimilation problems in science and engineering \cite{karniadakis2021physics}, such as subsurface transport in porous media \cite{he2020physics} , acoustic wave propagation \cite{rasht2021physics}, solid mechanics \cite{rao2021physics} and heat transfer \cite{cai2021physics}. 
In particular, PINN demonstrates tremendous success in simulating flow dynamics \cite{cai2022physics}, including high-speed aerodynamic flows \cite{mao2020physics}, incompressible flow governed by Navier Stokes equations \cite{sun2020surrogate,jin2021nsfnets},  laminar flows at low Reynolds numbers \cite{rao2020physics}, and even fluid-structure interaction \cite{raissi2019deep}. Previous applications of PINN to the hydraulic modeling \cite{cedillo2022exploratory}, despite limited, show that it can outperform Artificial Neural Network (ANN) in flood simulations \cite{mahesh2022physics}.

The ML-based downscaling is essentially a data assimilation problem with the ML models merging measurements and numerical model outputs into the ML training.
A key feature of PINN is its simplicity for assimilating observations \cite{he2020physics,he2021physics}. 
The time-varying observations and/or spatial snapshots at various spatiotemporal scales can be readily incorporated into the PINN training. 
Based on this feature, we propose a data assimilation approach based on the standard PINN framework and its variant to address problems in Fourier spaces (ff-PINN) \cite{wang2021eigenvector}.   
The proposed PINN computes the subgrid solution of a coarse model output and assimilated observations without modifying the numerical algorithms or refining the mesh resolution. 
By incorporating the dynamics presented in the measurements, PINN is able to emulate the nonlinear interactions of storm surge, tide and discharge within tidal rivers.
We do not intend to replace the numerical solver by PINN. Instead, we aim to build the PINN method upon the infrastructure of existing large-scale models to provide downscaled solutions at cells of interest.

The PINN-based method has a few additional advantages over local-scale numerical models. PINN is mesh-free, making it an efficient and flexible tool for downscaling. 
In practice, the downscaled solution at regions of interest may be obtained by configuring a local numerical model or an emulator at fine scales. Even though PINN itself may not be as efficient as the local model in terms of running speed, PINN does not require discretization and thus saves efforts of generating meshes for the numerical configuration. 
The meshless formalism enables flexible implementation in various domains as the partial derivatives are evaluated using automatic differentiation in Tensorflow \cite{abadi2016tensorflow}. 
Moreover, the PINN-based data assimilation allows exploring the potentials of merging various types of measurements from hydroacoustic meters, remote sensing platforms and other data collection instruments.

This research aims at developing a downscaling approach based on the PINN data assimilation to resolve subgrid variability of river dynamics in coastal regions for large-scale river models. 
A PINN solver of 1-D SVE is developed to merge in-situ and remote sensing measurements as well as coarse-scale model solutions to obtain a more accurate downscaled solution near the coastal interface and improve flood simulation. To the best of our knowledge, this is the first study in literature which develops a data assimilation model for downscaled river flow characterized by SVE with the utilization of physics-informed machine learning methods.
Several synthetic cases are tested to demonstrate our model's capability of reproducing flood propagation in a theoretical setup. Importantly, we develop a new neural network architecture to account for periodic tidal oscillations at the river downstream boundary and investigate the effects of the proposed network architecture on the PINN performance.
This manuscript is organized as follows. In Section \ref{s:methodology}, we present the PINN method for solving SVE and data assimilation. The PINN-based downscaling is also discussed. The problem formulation of each case study and the corresponding results are provided in Section \ref{s:setup}. Section \ref{s:discussion} presents the result discussions, potential limitations in realistic applications and the future path. Finally, the conclusions are provided in Section \ref{s:conclusion}.

\section{Methodology}
\label{s:methodology}

\subsection{Saint-Venant Equations}
\label{s:SVE}

The 1-D SVE consists of a continuity equation and a momentum equation for the dynamics of velocity ($u$) and water depth ($h$) along the river channel. The incompressible continuity equation is defined as
\begin{equation}
	\label{eq:continuity}
	\frac{\partial h}{\partial t} + u\frac{\partial h}{\partial x} = q,
\end{equation}
where $x$ is distance along the river channel, $t$ is time and $q$ is the water inflow per unit length of the channel from land surface and subsurface runoff, groundwater and precipitation. Throughout the study, it is assumed that the river channel does not receive water from the land and atmosphere ($q=0$). 
The momentum equation in the full dynamic form is
\begin{equation}
	\label{eq:momentum}
	\frac{\partial u}{\partial t} + u \frac{\partial u}{\partial x} + g \frac{\partial h}{\partial x} + g(S_f-S) = 0,
\end{equation}
where $g$ is gravity, $S$ is riverbed slope and $S_f$ is the friction slope that can calculated using the Chezy–Manning equation:
\begin{equation}
	\label{eq:Sf}
	S_f = \frac{n^2 |u|u}{R^{\frac{4}{3}}},
\end{equation}
in which Manning's roughness coefficient $n$ is used as the frictional coefficient and $R$ is the hydraulic radius. This study only considers rectangular channels, so $R$ can be expressed as 
\begin{equation}
	\label{eq:RH}
	R = bh/(2h + b),
\end{equation}
where $b$ is the channel width. 
The large-scale river models usually use simplified forms of the momentum equation, including the local inertial equation that neglects the convective acceleration term ($u \frac{\partial u}{\partial x}$) \cite{yamazaki2012analysis,de2013applicability}, the diffusive wave equation that neglects both the convective acceleration term and the local acceleration term ($\frac{\partial u}{\partial t}$) \cite{luo2017modeling}, and the kinematic wave equation that neglects all partial derivative terms \cite{li2013physically}. None of these simplified schemes include advection that dominates momentum propagation when the flood wave dynamics is strong \cite{dykstra2020propagation}.

\subsection{PINN approximation of SVE}
\label{s:PINN}

\begin{figure}[htbp]
	\centering
	\includegraphics[width=0.8\textwidth,keepaspectratio=true]{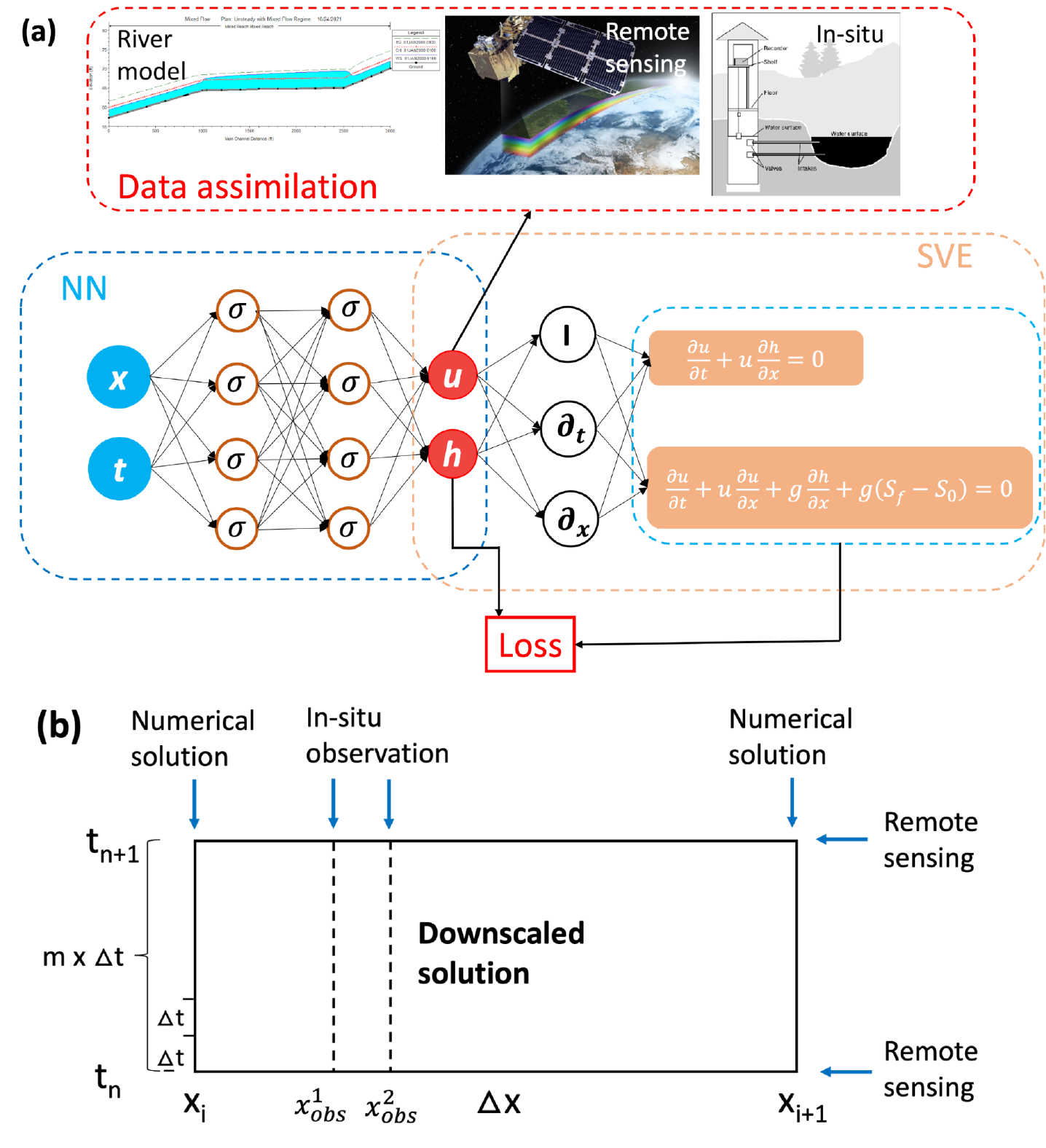}
	\caption{(a) The schematic of the PINN-based data assimilation model. The left part is the densely connected neural network with the input coordinates of $x$ and $t$ and output of $u$ and $h$. For illustration purposes, only 2 hidden layers and 4 neurons per hidden layer are shown. The right part is the SVE formulation and the operators computed by automatic differentiation, where $I$ denotes the identity operator. The upper part shows the data from numerical river models, remote sensing and in-situ measurements that can be assimilated in the PINN framework. (b) The Schematic of the PINN-based downscaled solution within a sub-grid cell of a large-scale river model over multiple time steps. The time step is denoted as $\Delta t$. The grid size is denoted by $\Delta x$, $x_i$ and $x_{i+1}$ are the cell's nodes, $t_n$ and $t_{n+1}$ are the time steps when the remote sensing snapshots are available, and $x_{obs}^1$ and $x_{obs}^2$ represent the locations of in-situ observations. }
	\label{fig:nn}
\end{figure}

In this study, instead of using numerical discretization, we take the first attempt to solve SVE using the PINN method. As illustrated in the schematic diagram (Figure \ref{fig:nn}(a)), 
the fully connected feed-forward deep neural network (DNN) in PINN takes spatial and temporal coordinates $x$ and $t$ as inputs and predicts the corresponding unknown variables $u$ and $h$ in the output layer. 
There are $l$ hidden layers between the input and output layers and $N_l$ neurons in each hidden layer. The neurons between the adjacent layers are fully connected and the inputs of the $l_{th}$ layer ($\bm{z}_l$) are fed from the outputs of the previous layer ($\bm{z}_{l-1}$): 
\begin{eqnarray}
	\label{eq:NN}
	\bm{z}_l = \sigma(\bm{W}_l\bm{z}_{l-1}+\bm{b}_l),
\end{eqnarray}
where the hyper-parameters $\bm{W}_l$ and $\bm{b}_l$ are the weight matrix and bias vector at the $l_{th}$ layer, determined after training, and $\sigma$ is the activation function used to introduce nonlinearity to each output component. In this study, the activation function is selected as $tanh(x)$, and $\bm{W}$ and $\bm{b}$ are initialized using a widely-used initialization scheme, Xavier scheme \cite{glorot2010understanding}, where initial weights are sampled from a truncated normal distribution.

The partial derivatives in the governing equations are used with the solutions predicted by DNN to approximate the residuals of the governing equations. The partial derivatives are obtained using the automatic differentiation, implemented in the deep learning platform Tensorflow \cite{abadi2016tensorflow} that computes the gradient of an output variable with respect to the input coordinates \cite{baydin2018automatic}. 

In PINN, the solutions of SVE, the instantaneous velocity $u(x,t)$ and water depth $h(x,t)$, are estimated with:
\begin{eqnarray}
	\label{eq:solution}
	& u(\bm{x},t) \approx \hat{u}(\bm{x},t,\theta), \\
	& h(\bm{x},t) \approx \hat{h}(\bm{x},t,\theta),
\end{eqnarray}
where $x$ and $t$ are the space and time vectors, and $\theta$ is the vector of weights and biases \cite{bishop2006pattern}. 

The loss function consists of the residuals of PDE (i.e. the summed residual errors of equation \ref{eq:continuity} and \ref{eq:momentum} when $\hat{u}$ and $\hat{h}$ are substituted), and the mean square errors of the DNN approximation against the boundary condition (BC) and the snapshot data associated with $u$ and $h$ (the snapshot data is denoted with symbol $S$):
\begin{equation}
	\label{eq:total_loss}
	J(\theta) = J_f(\theta) + \sum_j w_j J_j(\theta). 
\end{equation}
The subscript $j$ is $BC_{u}$, $BC_{h}$, $S_{u}$ and $S_{h}$, $J$ and $w$ represent the loss terms and weighting coefficients correspondingly. 
The PDE loss ($J_f$) is defined as
\begin{equation}
	\label{eq:J_f}
	J_f(\theta) = \frac{1}{N_f} \sum^{N_f}_{i=1} | {r_f}^2(\bm{x}^i,t^i,\theta) |,
\end{equation}
where $N_f$ denotes the number of collocation points in the computational domain. 
The residual of SVE ($r_f$) is obtained by summing the continuity equation \eqref{eq:continuity} and the momentum equation \eqref{eq:momentum} with $\hat{u}$ and $\hat{h}$ substituted.  The loss functions corresponding to BC and spatial snapshot data are defined as:
\begin{eqnarray}
	\label{eq:BC_loss}
		& J_{BC_u}(\theta) & = \frac{1}{N_{BC_u}} \sum^{N_{BC_u}}_{i=1} | \hat{u}(\bm{x}^i,t^i,\theta) - u(\bm{x}^i,t^i) |^2 , \nonumber \\
		& J_{BC_h}(\theta) & = \frac{1}{N_{BC_h}} \sum^{N_{BC_h}}_{i=1} | \hat{h}(\bm{x}^i,t^i,\theta) - h(\bm{x}^i,t^i) |^2 , \quad \bm{x} \in \Omega_{BC}, t\in [0,T]
\end{eqnarray}
\begin{eqnarray}
	\label{eq:S_loss}
		& J_{S_u}(\theta) & = \frac{1}{N_{S_u}} \sum^{N_{S_u}}_{i=1} | \hat{u}(\bm{x}^i,0,\theta) - u(\bm{x}^i,0) |^2 , \nonumber \\
		& J_{S_h}(\theta) & = \frac{1}{N_{S_u}} \sum^{N_{S_u}}_{i=1} | \hat{h}(\bm{x}^i,0,\theta) - h(\bm{x}^i,0) |^2 , \quad \bm{x} \in \Omega_{S}
\end{eqnarray}
where $N_{BC}$ refers to the number of points sampled at the boundaries of the computational domain and $N_{S}$ is the number of points sampled at the spatial snapshots, and $T$ is the simulation period.
Time-varying data, such as the observations of $u$ and $h$, can be used in the PINN training to add further constraints. In this study, we add two additional terms (i.e. $w_{obs_{u}} J_{obs_{u}}$ and $w_{obs_{h}} J_{obs_{h}}$) to the total loss in equation \ref{eq:total_loss}. The loss functions corresponding to the observations are
\begin{eqnarray}
	\label{eq:obs_loss}
		& J_{obs_u}(\theta) & = \frac{1}{N_{obs_u}} \sum^{N_{obs_u}}_{i=1} | \hat{u}(\bm{x}^i,t^i,\theta) - u(\bm{x}^i,t^i) |^2 , \nonumber \\
		& J_{obs_h}(\theta) & = \frac{1}{N_{obs_h}} \sum^{N_{obs_h}}_{i=1} | \hat{h}(\bm{x}^i,t^i,\theta) - h(\bm{x}^i,t^i) |^2 , \quad \bm{x} \in \Omega_{obs}, t\in [0,T]
\end{eqnarray}

Previous studies showed that the weighting coefficients of the loss functions ($w$) in equation \eqref{eq:total_loss} are critical to the PINN training as these coefficients are used to balance the contribution of different loss terms \cite{wang2021eigenvector,van2022optimally,2020AGUFMH036.0003H,he2021physics}. 
Imbalanced loss terms can lead to failure in PINN. The selection of weights is problem-specific because the optimal combination of weights varies across different flow scenarios, depending on system properties and conditions. The weights are typically tuned arbitrarily before training via a trial-and-error procedure to search for the optimal solution, which is also referred as nonadaptive weighting, but such method can be very time-consuming \cite{jin2021nsfnets}.
Recently, a self-adaptive weight approach for PINN has been proposed in \cite{mcclenny2020self} that trains the weights by autonomously focusing on the difficult regions of the solution. The method attains satisfactory accuracy but significantly increases the computational cost in training. 
To address this issue, in this study we adopt the strategy of adaptive weights \cite{wang2021understanding} that scales different loss terms by updating the coefficients to balance gradients during the back-propagation of the DNN training. The weights at every iteration of the training are updated based on the loss ($J$) obtained in the previous iteration. This scheme is demonstrated efficient and effective in our numerical study. The formulation of the DNN parameters at each iterative step is 
\begin{equation}
	\label{eq:parameter_update}
	\theta_{k+1} = \theta_k - \eta | \nabla_{\theta}J_f(\theta_k) + \sum_j \nabla_{\theta} J_j(\theta_k) |, 
\end{equation}
where $k$ is the iteration step and $\eta$ is the learning rate. The estimated weighting coefficients ($\hat{w_j} $) at the ($k+1$)th iteration are computed as
\begin{equation}
	\label{eq:weight_update}
	\hat{w_j} = \frac{max_\theta{|\nabla_{\theta}J_f(\theta_n)|}}{\bar{\nabla_{\theta} J_j(\theta_n)}}. 
\end{equation}
The new weighting coefficients are then updated with a moving average
\begin{equation}
	\label{eq:weight_new}
	w_j = (1-\alpha)w_j + \alpha \hat{w_j}. 
\end{equation}
The hyper-parameter $\alpha$ determines the decay rate of the previous weight and is taken as 0.9 to ensure a stable adjustment during the training \cite{jin2021nsfnets,wang2021understanding}. 

In the PINN training, the DNN is trained using the stochastic gradient descent method by minimizing the loss function of equation \ref{eq:total_loss} using an adaptive optimization algorithm, Adam optimizer \cite{kingma2014adam}. 
The learning rate is a hyper-parameter that determines how much $\theta$ is updated at each iteration step of the training process. Here we apply the learning rate annealing algorithm with an initial learning rate and an exponential decay both of 0.9 \cite{wang2021understanding,rasht2021physics}. 
Normalization of the input coordinates is important to ensure the convergence to the correct solution. So the space and time are both mapped to $[-1,1]$:
\begin{equation}
	\label{eq:normalization}
	\bm{X} = \frac{2\bm{X}-min(\bm{X})}{max(\bm{X})-min(\bm{X})} - 1,  
\end{equation}
where $\bm{X}$ represents the input vector of $\bm{x}$ and $t$. 

\subsection{Fourier feature embedding}
\label{s:ff}

In practice, the training of a PINN model is challenging. The PINN solutions may be trapped to a local minimum (i.e., a trivial solution), which satisfies the PDE residual loss but is far from the true solution that should also satisfy initial and boundary conditions \cite{wong2021learning}. This is because the trivial solution of a PDE results in flat outputs under the activation function, e.g., $tanh$. This issue can be mitigated by transforming the inputs of PINN into a multi-scale feature space, such as through Fourier feature mapping \cite{tancik2020fourier,wang2021eigenvector} or the sinusoidal mapping \cite{wong2021learning}. Such transformation can modulate the levels of the output variability to match the high-frequency patterns of emulated physics in time and space. 

As proposed by \cite{wang2021eigenvector}, prior to the fully connected neural network, we introduce a multi-scale Fourier feature architecture to apply the Fourier feature embeddings to input coordinates:
\begin{equation}
\label{eq:fourier_mapping}
\bm{\gamma}(\bm{\nu}) = 
\begin{bmatrix}
	cos(2\pi \bm{B} \bm{\nu})\\
	sin(2\pi \bm{B} \bm{\nu})
\end{bmatrix},
\end{equation}
where $\gamma$ is the Fourier feature mapping function, $\bm{\nu}$ is the input coordinates, $\nu=[\bm{x},t]^T$, and $\bm{B}$ is the mapping matrix corresponding to the frequency of the multi-scale Fourier features. The values in $\bm{B}$ are sampled from a Gaussian distribution $N(0, \textit{s})$. The hyper parameter $\textit{s}$ is referred as bandwidth in \cite{wong2021learning}, which is related with the frequency range of the Fourier features and directly affects the output variability. This parameter is problem-dependent and needs to be adjusted to match the frequency of the problem. The value of $\textit{s}$ can either be optimized during the PINN training or manually fixed based on the prior knowledge of the boundary data.
With the Fourier feature embeddings, the first layer of the neural network becomes
\begin{equation}
\label{eq:layer1}
\bm{H}_1 = \phi(\bm{W}_1 \cdot \bm{\gamma} (\bm{\nu})+\bm{b}_1),
\end{equation}
where $\phi$ denotes the activation function, $W_1$ and $b_1$ represent the weights and biases, respectively, in the first hidden layer.  

In this study, we devise a new neural network architecture that only applies the Fourier feature embeddings to the temporal coordinate to account for the periodic feature of the tidal boundary forcing. As Fourier feature embedding provides the initial guess to the target solutions, the physical characteristics, such as the periodic tidal patterns, are seamlessly encoded into PINN. 
At the downstream end of a river, the dynamic tidal wave generated by the ocean tides and storm surge can propagate far upstream into the river network \cite{ikeuchi2017compound}, inducing high-frequency oscillations in flow velocity and water depth. We address the tidal wave propagation over multiple tidal cycles by mapping the time into a multi-scale Fourier space. 
The feed forward pass for time and space is now defined as
\begin{equation}
\label{eq:mapping_time}
\bm{\gamma}_t^{(i)}(t) = 
\begin{bmatrix}
	cos(2\pi \bm{B_t}^{(i)} t)\\
	sin(2\pi \bm{B_t}^{(i)} t)
\end{bmatrix}, \quad \bm{H}_{t,1}^{(i)} = \phi(\bm{W}_1 \cdot \bm{\gamma}_t^{(i)}(t) + \bm{b}_1), \quad for \; i = 1, 2, ...
\end{equation}
\begin{equation}
\label{eq:mapping_space}
\quad \bm{H}_{x,1} = \phi(\bm{W}_1 \cdot \bm{x} + \bm{b}_1),
\end{equation}
where $i$ represents the index of the spatial dimension in the Fourier space. In a multi-scale problem, it is possible to map the time input to several Fourier features with different $\textit{s}$ to reproduce variability sources of different frequencies. The choice of $\textit{s}$ is problem-dependent and $\textit{s}$ usually falls between 10$^{-1}$ and 10$^2$. 
The value of $\textit{s}$ may be determined by trial and error or performing spectral analysis on the training data \cite{wang2021eigenvector}. The parameter may also be treated as a trainable hyper-parameter and tuned in the PINN training by monitoring the loss function \cite{wong2021learning}.

\subsection{The PINN-based downscaling}
\label{s:subgrid}

This section introduces the PINN-based downscaling method that provides an efficient subgrid solution of a large-scale river model. In such cases, the numerical solution of the coarse-scale model, remote sensing data and in-situ measurements provide the BC, spatial snapshots and observations, respectively, required for the PINN training and validation. 

Figure \ref{fig:nn}(b) shows the diagram of using PINN to produce a downscaled solution within a grid cell of a 1-D numerical model over multiple time steps. 
The cell of interest with nodes $x_i$ and $x_{i+1}$ is the domain of a PINN model. The numerical solutions at the grid nodes are used as the boundary data for the PINN training, which are provided at the time interval $\Delta t$ that is usually set as less than one second to hundreds of seconds in river modeling. 

The spatial snapshots overlapping the grid cell ($x\in[x_i,x_{i+1}]$) can be extracted from high-resolution remote sensing data whenever available, e.g., at time steps $t_n$, $t_{n+1}$. 
The remote sensing snapshots usually have a much higher spatial resolution (O($\sim$100m)) than the large-scale river model (O($>$1000m)) and a much lower temporal resolution ($\sim$ several day) than the model's output frequency (e.g. 1 hour). 
Within the river model's grid cell, since PINN does not require grid discretization, the collocation points are directly made from the remote sensing pixels. In the case when remote sensing data are unavailable at the exact location of the model's grid node, the nearest data will be used.
It is worth noting that it is difficult to estimate flow velocity accurately from remote sensing. To mimic this data limitation, we only consider the water depth information in the spatial snapshots and focus on the downscaled solution of water depth that is of most significance to riverine/coastal flooding.

In-situ measurements of velocity and water depth exist extensively in river systems, particularly coastal river reaches where tidal gauges are installed. For example, United States Geological Survey (USGS) maintains over 8200 continuous records of streamgages for over 115 years \cite{turnipseed2010discharge}. The water stage is continuously recorded in USGS gauges usually every 15 minutes with less than 0.2\% error. The instruments such as current meter and Acoustic Doppler Current Profiler (ADCP) are used to measure flow velocity periodically, which is then used to derive and maintain the semi-empirical stage-discharge relation for continuous velocity estimates. Many other state agencies and private sectors are also consistently contributing to in-situ measurements.
The in-situ measurements usually have continuous coverage over the simulation period ($t \in [t_n, t_{n+1}]$) and small time interval, usually from 6 minutes to 1 hour. These time-series data wherever available (e.g. at $x_{obs}^1$, $x_{obs}^2$) are readily assimilated into PINN. Table \ref{table:esm} summarizes the information of the assimilated data used in the PINN training.
We will demonstrate the effectiveness of using PINN to solve the downscaled solutions in case 5 and 6 of the following section.

\begin{table}[htbp]
	\footnotesize
	\caption{Information of PINN training data.}
	\centering
	\begin{tabular}{ p{3.5cm} p{1.8cm} p{2.4cm} p{2.0cm} p{2.5cm}}
		\hline
		Data source & Loss function & Location & Period & Time interval \\
		\hline
		  Large-scale model & $J_{BC}$ & $x=x_i$, $x_{i+1}$ & $t \in [t_n, t_{n+1}]$ & 1 hour  \\
		Remote sensing snapshots   & $J_{S}$ & $x \in [x_i, x_{i+1}]$ & $t=t_n, t_{n+1}$, ... & $\sim$10 day \\	
		In-situ measurements   & $J_{obs}$ &  $x=x_{obs}^{1}$, $x_{obs}^{2}$,... $x_{obs}\in [x_i, x_{i+1}]$ & $t \in [t_n, t_{n+1}]$ & 6 minutes to 1 hour \\
		\hline
	\end{tabular}
	\label{table:esm}
\end{table}

\section{Problem Formulation}
\label{s:setup}

In this section, we used six synthetic experimental cases, building up in complexity, to explore the feasibility of using PINN to downscale a large-scale river model for tidal rivers.  
The flow simulations are performed over an initially-dry floodplain and along an idealized open channel forced by upstream discharge and downstream tide and storm surge. 
We first evaluate the capability of the PINN data assimilation in resolving flow dynamics (Case 1$\sim$3) and that of the Fourier feature embedding scheme in capturing the high-frequency tidal variations (Case 4). The PINN solutions are compared with analytical or numerical solutions. We then compare the PINN's performance in downscaling the spatially-varied channel flow with that of the traditional linear interpolation method (Case 5$\sim$6).

The first two test cases in section \ref{s:case1} and section \ref{s:case2} follow the numerical experiments performed in \cite{hunter2005adaptive,de2013applicability}, which simulated flood wave propagation over an initially-dry plane with zero and uniform adverse slope, respectively.  
In these cases, the velocity ($u$) is assumed to be constant. The SVE is simplified and the water depth ($h$) is the single unknown variable. Thus, both the analytical solution and a 4th-order Runge-Kutta (RK4) solution can be used to calculate $h$.  
The third and fourth test cases in section  \ref{s:case3} and section \ref{s:case4} simulate flood propagation induced by a pseudo storm surge and a combination of tide and surge, respectively, in a tilted open channel.  Both $u$ and $h$ are unknown. The reference solutions of $u$ and $h$ are simulated by a high-fidelity hydraulic model. The first four cases are used to demonstrate the capability of PINN in solving 1-D SVE at unsteady flow conditions, which is required to accurately downscale river models in coastal regions.
Finally, in case 5 and 6, we explore to use PINN to solve the downscaled solutions with changing river topology and flow regimes. The size of NN increases with the problem size and the number of unknown variables \cite{raissi2019deep}. In this study, we use 32 neurons for each state variable and manually tune the number of hidden layers in each test case. The PINN configurations and numerical setups are summarized in Table \ref{table:experiment}.

Among all cases, the workflow is organized as:
\begin{itemize}
  \item Set up the analytical or numerical models in the idealized domain to simulate the reference solutions of $u$ and/or $h$. The experimental configurations are summarized in Table \ref{table:experiment}. 
  \item Prepare the training data. While the boundary data ($J_{BC_u}$) and ($J_{BC_h}$) are directly obtained from the reference solutions, the pseudo observations, such as spatial snapshots ($J_{obs_u}$ and $J_{obs_h}$) and in-situ measurements ($J_{S_h}$), are estimated from the reference solutions by adding a small level (0.2\%) of random white noise. This noise level is determined as the typical error in the USGS water depth measurements \cite{turnipseed2010discharge}. 
  \item Formulate the PINN scheme for each test case with manually-tuned NN layers and number of neurons per layer (Table \ref{table:experiment}). 
  \item Train the PINN model as outlined in Figure \ref{fig:nn}(a). 
  \item Evaluate the PINN solutions against the reference solutions (Case 1$\sim$6) and the linear-interpolated downscaled solutions (Case 5 and 6) using two model skill metrics: the relative $L_2$ error 
\begin{equation}
\label{eq:L2}
\epsilon_h = \sqrt{ \frac{\sum_{i=1}^N (\hat{h_i}-h_i)^2}{\sum_{i=1}^N {h_i}^2} },
\end{equation}
and root mean squared error (RMSE)
\begin{equation}
\label{eq:rmse}
\text{RMSE} = \sqrt{\frac{1}{N}\sum_{i=1}^N (\hat{h_i}-h_i)^2},
\end{equation} 
where $\hat{h}$ and $h$ are the predicted and referenced water depth, respectively, and $N$ is the number of data points. 
\end{itemize}

\begin{table}[htbp]
	\caption{Experimental configurations. }
	\footnotesize
	\centering
	\begin{tabular}{ p{0.8cm} | p{1.2cm} p{1.2cm} p{1.8cm} p{1.2cm} | p{1.8cm} p{1.2cm} | p{1.2cm} p{1.2cm}} 
		\hline
		      &  \multicolumn{4}{c|}{Numerical setup}  & \multicolumn{2}{c|}{Pseudo observations} & \multicolumn{2}{c}{NN structure} \\\cline{2-9}
                     & domain & simulation period & downstream BC & upstream BC  & gauge & snapshot & layers & neurons per layer \\
		\hline
		case 1   & 0-3600 m& 0-3600 s& eq.\eqref{eq:case3_upstrm_bnd}& h=0 at x=ut& x=1200, 2400 m & t=3600 s& 3& 32\\
            \hline
		case2   & 0-3600 m& 0-3600 s& eq.\eqref{eq:BC}& h=0 at x=ut& x=1200, 2400 m & t=3600 s& 3& 32\\
            \hline
		case3   & 0-305 m& 0-24 h& eq.\eqref{eq:BC}& 5.7 m$^3$/s& x=100, 200 m & t=0 h& 3& 64\\
		\hline
            case4   & 0-305 m& 0-240 h& tide and surge (Fig. \ref{fig:case456_BC})& 5.7 m$^3$/s& x=100, 200 m & t=0 and 10 day& 4& 64\\
            \hline
            case5   & 0-914 m& 0-240 h& S=0.001& discharge (Fig. \ref{fig:case456_BC})& x=100, 200 m & t=0 and 240 h& 5& 64\\
            \hline
            case6   & 0-914 m& 0-240 h& tide and surge (Fig. \ref{fig:case456_BC})& discharge (Fig. \ref{fig:case456_BC})& x=100, 150, 200, 300, 600 m & t=0 and 240 h& 5& 64\\
            \hline
	\end{tabular}
	\label{table:experiment}
\end{table}

\subsection{Case 1: Nonbreaking wave propagation over a horizontal plane}
\label{s:case1}

The first case explores the PINN's capability to simulate the flood wave propagation over a horizontal plane, in which the inundation can be considered as an advancing front moving at the speed of $u$. The analytical solution of SVE is available by assuming $u$ to be constant over space and time \cite{hunter2005adaptive}. 
Thus, equations \eqref{eq:continuity} and \eqref{eq:momentum} are rewritten as
\begin{equation}
	\label{eq:case1_continuity}
	\frac{\partial h}{\partial t} + u \frac{\partial h}{\partial x} = 0
\end{equation}
\begin{equation}
	\label{eq:case1_momentum}
	\frac{\partial h}{\partial x} = -(S + \frac{n^2u^2}{h^{4/3}}). 
\end{equation}
At $x=ut$, the moving boundary condition is $h(ut,t)=0$. As $S=0$, the analytical solution of $h$ can be obtained by directly integrating equation and imposing the moving boundary, which yields 
\begin{equation}
	\label{eq:case3_solution}
	h(x,t) = {-\frac{7}{3}[n^2u^2(x-ut)]}^{3/7}. 
\end{equation}
At $x=0$, the boundary condition is derived from the analytical solution:
\begin{equation}
	\label{eq:case3_upstrm_bnd}
	h(0,t) = (\frac{7}{3}n^2u^3t)^{3/7}. 
\end{equation}
The reference solution \eqref{eq:case3_solution} is solved by setting $u$=1 m/s, $n$=0.005  m$^{-1/3}$s, $\Delta x$=30 m and $\Delta t$=30 s. The simulation period is 3600 s. 

The PINN solution associated with equations \eqref{eq:case1_continuity} and \eqref{eq:case1_momentum} is obtained using a DNN with 3 hidden layers and 32 neurons per hidden layer. At $t=0$, the inundation front has not yet moved. Only the water depth at $x=0$ is known at the initial state. So the spatial snapshot is only enforced at $t$=3600 s. The computational domain is restricted by the far end of the inundation front that represents the propagation distance it has moved over time $t$. Thus, the BC derived from equation \eqref{eq:case3_upstrm_bnd} is enforced at $x=0$, and $h=0$ is imposed at the moving boundary ($x=ut$) over the simulation, $t\in[0,3600]$. Note that it is straightforward to impose moving boundary condition under the PINN framework owing to its space-time data assimilation capacity, see equation \eqref{eq:obs_loss}. The resulting $N_{S}$ and $N_{BC}$ are 121 and 242, respectively. Although PINN does not require mesh generation, the collocation points to compute the loss functions are specified at the nodes of the discretized mesh for the comparison with the analytical solution or the numerical solutions in the following sections. The learning rate is specified with an initial value of 10$^{-4}$ that is reduced exponentially every 5000 iterations at the rate of 0.9. The in-situ observations are assimilated at the point of $x$=1200 m and 2400 m over the entire simulation period ($N_{obs}=242$). 

\subsection{Case 2: Nonbreaking wave propagation over an adverse slope}
\label{s:case2}

In the second test case we simulate the flood propagation over an adverse slope ($S \neq 0$). While the equations \ref{eq:case1_continuity} and \ref{eq:case1_momentum} are still valid, an analytical solution is no longer available. As a result, the numerical solution of $h$ is solved at each time step using the RK4 scheme \cite{hunter2005adaptive,de2013applicability}:
\begin{equation}
\label{eq:RK4}
h_{i+1} = h_{i} +\frac{1}{6} \Delta x (k_1 + 2k_2 + 2k_3 +k_4),	
\end{equation}
where 
\begin{eqnarray}
\label{eq:ks}
& k_1 & = f(h_i), \nonumber \\
& k_2 & = f(h_i+\Delta x\frac{k_1}{2}), \nonumber \\
& k_3 & = f(h_i+\Delta x\frac{k_2}{2}), \nonumber \\
& k_4 & = f(h_i+\Delta xk_3),
\end{eqnarray}
where 
\begin{equation}
f(h_i) = - (S + \frac{n^2u^2}{{h_i}^{4/3}} ). 
\label{eq:f}
\end{equation}
Since there is no analytical boundary condition, we use a sinusoidal wave to emulate an advancing inundation front on the floodplain:

\begin{equation}
	\label{eq:BC}
	h(0,t) = A \sin(2\pi t/T_{BC}).
\end{equation}
The amplitude $A$ is 4 m and the period $T_{BC}$ is 4 h. 
Please note that the Manning's $n$ determines the maximum travel distance, beyond which the floodplain is dry. However, considering the maximum distance that the flood wave can travel is $x=ut$, we can impose a vanishing condition at the moving boundary $x=ut$, that is $h(ut,t)=0$. 
In the simulation, the Manning's $n$ is specified as 0.025 m$^{-1/3}$s and $S$ is $10^{-3}$ m$\cdot$m$^{-1}$. The simulation period is 3600 s. The RK4 solution is obtained with setting $\Delta x=20$ m and $\Delta t=30$ s. The DNN is configured the same as that in section \ref{s:case1}. We evaluated the PINN solutions with assimilating observations at $x$=1200 m and 2400 m. The values of $N_{S}$, $N_{BC}$ and $N_{obs}$ are the same as those in case 1.

\subsection{Case 3: Storm surge propagation along a sloping open channel}
\label{s:case3}

This case examines the PINN's capability of solving the full SVE in a more dynamic flow condition. Because both analytical solution and RK4 solution are no longer available, the reference solution is obtained from a widely used finite difference based hydraulic model, HEC-RAS \cite{brunner1995hec}. The numerical solutions from HEC-RAS provide the BC, pseudo spatial snapshots and in-situ observational data for the PINN training. The convergence criteria/tolerance is set as 0.006 m (0.02 ft).

\begin{figure}[htbp]
	\centering
	\includegraphics[width=5in,keepaspectratio=true]{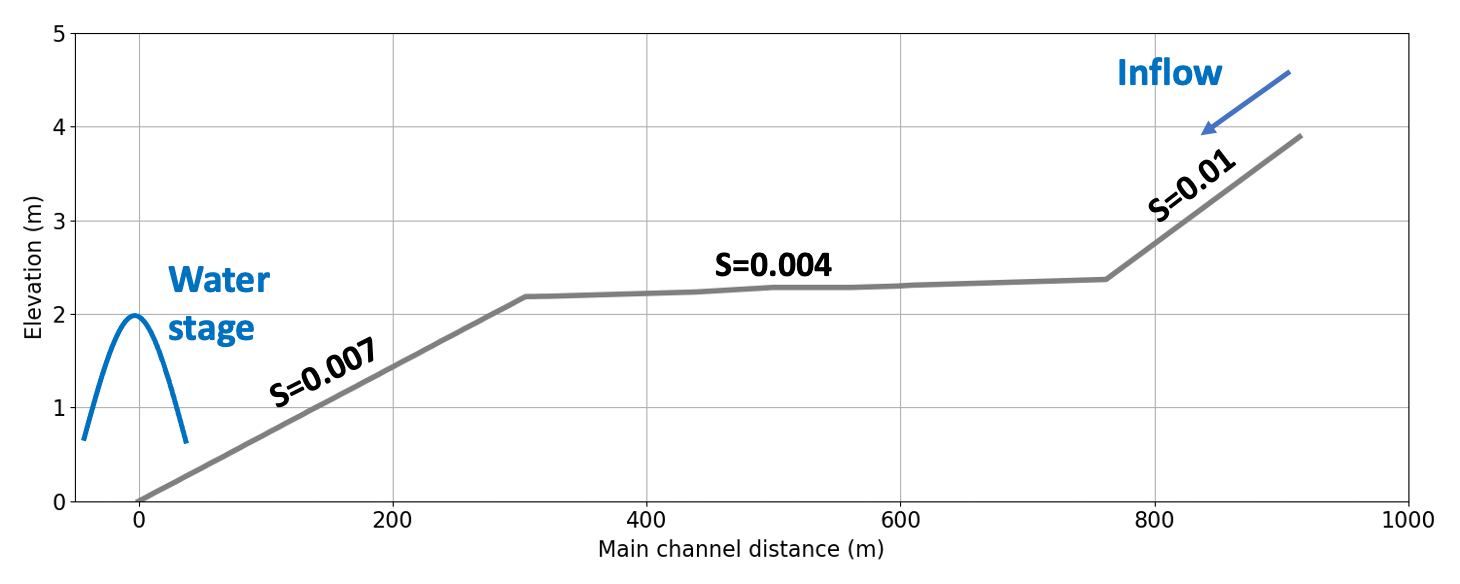}
	\caption{The diagram of the computational domain.}
	\label{fig:channel}
\end{figure}

The experiment is set up in an open channel with a uniform width of 3 m (Figure \ref{fig:channel}). The slope is approximately 0.007, 0.004 and 0.01 m m$^{-1}$ over the three sections. The flow regime changes with the channel slope. In HEC-RAS, the channel is divided into segments connected at nodes and the spatial resolution is 6 m. The state variables ($h$ and $u$) are outputted every 15 min.  
A constant flow of 5.7 m$^3$/s (200 cfs) is imposed at the upstream boundary, $x=914$ m (3000 ft),  while a pseudo storm surge is enforced at the downstream boundary, $x=0$ m. The storm surge is approximated using a periodic sinusoidal wave (equation \ref{eq:BC}) with an 2.4-m amplitude and a 50-h period. The Manning's $n$ is 0.022 m$^{-1/3}$s. The simulation period is 24 hours. Please note that US customary Units used in HEC-RAS simulations are converted to standard international units.

The DNN architecture consists of 3 hidden layers and 64 neurons per hidden layer. The collocation points are specified at each computational node. The learning rate is initialized at 5$\times$10$^{-4}$ and then reduced exponentially every 5000 iterations at the rate of 0.9. 
For case 3 and 4, we only focus on the flood propagation in the downstream section, $x=0\sim305$ m ($0\sim1000$ ft) where tide and storm surge dominate the momentum. The collocation points in the remaining domain are not used in the model training. 
The upstream and the downstream BC are imposed at $x=305$ m and $x=0$ m, respectively. The spatial snapshot of $h$ is only specified at $t=0$ as the revisit time of satellites is usually greater than a day. The time-series observations are assumed available every 15 min and are assimilated at $x$=100 m and 200 m. The resulting $N_{S}$, $N_{BC}$ and $N_{obs}$ are 51, 192 and 96, respectively.
The BC of $u$ and $h$ are obtained from the HEC-RAS output. The snapshot of $h$ and observational data of $u$ and $h$ are obtained by adding 0.2\% noise to the HEC-RAS solutions.

\subsection{Case 4: Tide and surge propagation along a sloping open channel}
\label{s:case4}

\begin{figure}[htbp]
	\centering
	\includegraphics[width=0.8\textwidth,keepaspectratio=true]{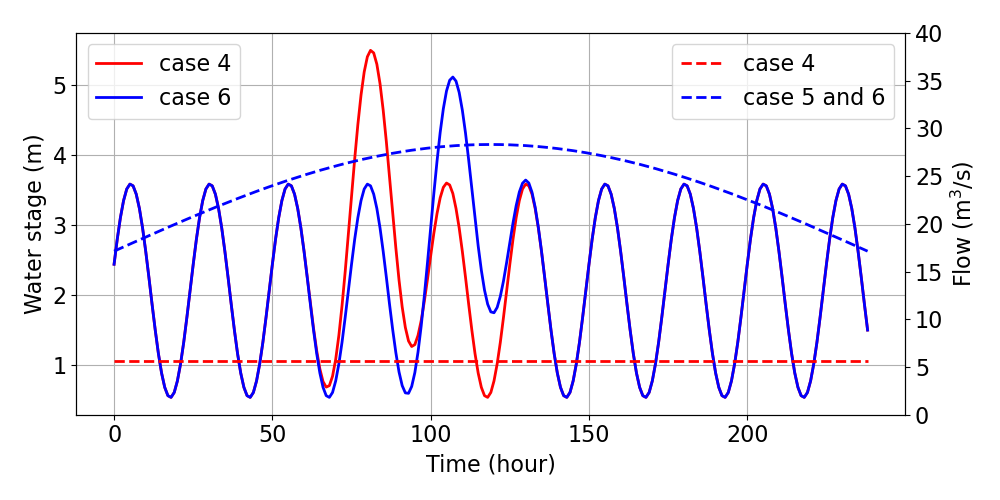}
	\caption{The upstream boundary of flow (dash lines) and the downstream boundary of water stage (solid lines) in case 4, 5 and 6.}
	\label{fig:case456_BC}
\end{figure}

In the test case 4, we simulated flood wave propagation forced by a combination of tide and storm surge along the same open channel as used in Section \ref{s:case3}, with a focus on examining the effect of Fourier feature embedding. The diurnal tidal signal is set by $A=1.5$ m and $T_{BC}=25$ h in equation \ref{eq:BC}. A storm surge is imposed to the tide during a randomly-selected 48-hour period. 
The upstream flow is 5.7 m$^3$/s. The simulation is extended to 10 days. The upstream and downstream boundaries specified in the HEC-RAS for the test case 4 are provided in Figure \ref{fig:case456_BC}. The state variables ($h$ and $u$) are outputted every 1 hour.  
The DNN architecture of 4 hidden layers is employed with each hidden layer consisted of 64 neurons. The other configurations are the same as those in section \ref{s:case3}, except that the snapshots are added at the initial state and the last time step assuming satellite data are available every 10 days. The corresponding $N_{S}$ is 102, and $N_{BC}$ and $N_{obs}$ are 480 and 240 according to the updated output interval.
We evaluate the PINN solutions under two different NN architectures: the standard NN architecture and the NN architecture using the Fourier feature embeddings. In both solutions, observations are added at $x$=100 and 200 m. 
We use $\textit{s}=$ of 0.25 and 10 (in Equation \ref{eq:mapping_time}) to account for the tidal frequency in this case. These numbers are determined by trial and error.

\subsection{Case 5: PINN downscaling}
\label{s:case5}

The aforementioned cases have shown that PINN can achieve reasonable accuracy when constrained by BC and data of spatial snapshots and in-situ observations. 
In case 5 and 6, we perform the simulation over the full channel (Figure \ref{fig:channel}) and explore the PINN's capability to resolve the channel topography and variable flow regimes at the subgrid scale. 
The PINN-based downscaled solutions are also compared with the solutions obtained from the linear interpolation. 

At the upstream boundary, a time-varying discharge is imposed with a base flow of 17 m$^3$/s (600 cfs) and a peak flow of 28.3 m$^3$/s (1000 cfs), which follows the sinusoidal curve (equation \ref{eq:BC}) with a period of 20 days and an amplitude of 11.3 m$^3$/s (400 cfs) as shown in Figure \ref{fig:case456_BC}. 
The downstream is configured with a normal depth boundary, where a constant slope ($S$) of 0.001 is added. This slope is slightly smaller than the slope of the downstream section and ensures that the water slowly drains. 

As the channel is considered as a grid cell of a large-scale river model, the downstream and upstream ends of the channel (at $x=$ 0 and 914 m) are the cell's nodes, corresponding to $x_i$ and $x_{i+1}$ in Figure \ref{fig:nn}(b), where the HEC-RAS solutions of $h$ and $u$ are enforced as the boundary condition in PINN. The pseudo in-situ observations over the simulation period are imposed at $x$=100 and 200 m. The pseudo snapshots of $h$ along the channel are added at $t=$0 and 240 h, corresponding to $t_n$ and $t_{n+1}$ in Figure \ref{fig:nn}(b). The linear-interpolation solution is also obtained by linearly interpolating water depth $h$ across the two boundary nodes and the in-situ observations at each time step.
A fully-connected DNN is employed with 5 hidden layers and 64 neurons per hidden layer. The other configurations are the same as those in section \ref{s:case4}. 
Since no tides are enforced in this case, PINN applies the standard NN architecture.

\subsection{Case 6: PINN downscaling with a periodic boundary}
\label{s:case6}

This case simulates a more dynamic condition at the river-ocean interface where oceanic and fluvial processes dominate the downstream and upstream sections separately. The impact of in-situ observations is  assessed by adding additional gauging stations to the domain. In addition to the two stations in the downstream section ($x$=100 and 200 m), we sequentially add gauges 150 m, 300 m and 600 m from the upstream boundary (Table \ref{table:case56}). This allows us to evaluate the performance of linear interpolation downscaling with more uniformly distributed in-situ observations. 

The upstream boundary is configured the same as that in case 5 (Section \ref{s:case5}).
The downstream water stage boundary, as shown in Figure \ref{fig:case456_BC}, is a combination of tide and storm surge as specified in Section \ref{s:case4}. The PINN solutions apply the NN architecture with the Fourier feature embeddings. The other configurations are the same as those in case 5.

\section{Result and Discussion}
\label{s:discussion}

This section demonstrates and discusses the results of the six test cases that are configured following the descriptions in Section \ref{s:setup}. The limitations of the PINN-based downscaling and future developments of the method are also discussed.

\subsection{Test results}

\subsubsection{Case 1 and 2}

\begin{figure}[htbp]
	\centering
	\makebox[\textwidth][c]{\includegraphics[width=\textwidth,keepaspectratio=true]{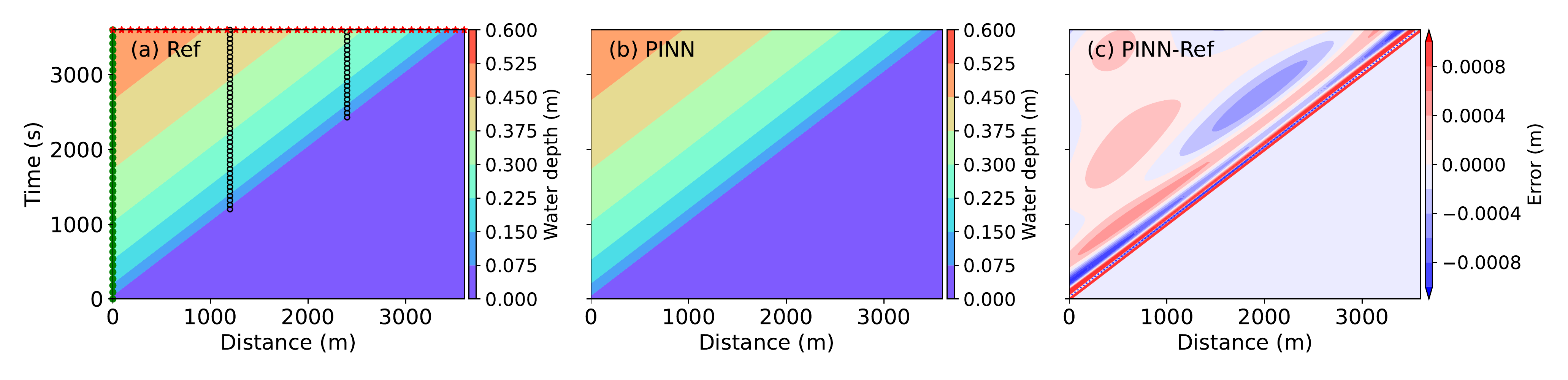}}
	\caption{The spatiotemporal evolution of the water depth in case 1 given by the analytical solution (a), the PINN solution (b) and the error in the PINN solution (c). The green dots, black circles and red stars in (a) of this figure and Figure \ref{fig:case2_contour}, \ref{fig:case3_contour}, \ref{fig:case4_contour}, \ref{fig:case5_contour} and \ref{fig:case6_contour} represent the training data locations of BC, in-situ observation and spatial snapshots, respectively.} 
	\label{fig:case1_contour}
\end{figure}

\begin{figure}[htbp]
	\centering
	\includegraphics[width=\textwidth,keepaspectratio=true]{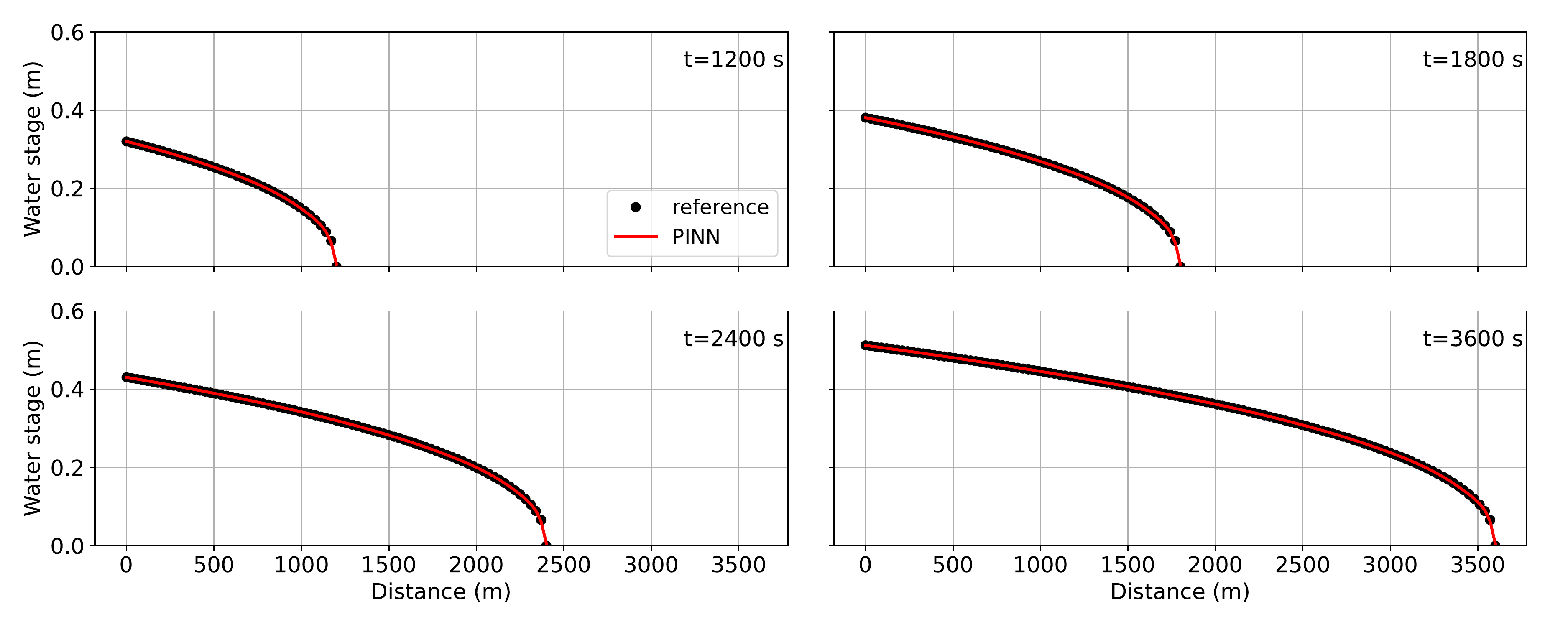}
	\caption{The water stage of the analytical solution and the PINN solutions at four time instants in case1.}
	\label{fig:case1_along_channel}
\end{figure}

We compare the spatiotemporal evolution of water depth ($h$) in the analytical solution and the PINN solutions (Figure \ref{fig:case1_contour}) along the flood propagation distance at four different time instants (Figure \ref{fig:case1_along_channel}). The result shows that the PINN solution achieves high accuracy. The pattern of the flood wave propagation and the shape of the inundation front are both well reproduced, with small $\epsilon_h$ and RMSE (Table \ref{table:case1234}). 

\begin{figure}[htbp]
	\centering
	\makebox[\textwidth][c]{\includegraphics[width=\textwidth,keepaspectratio=true]{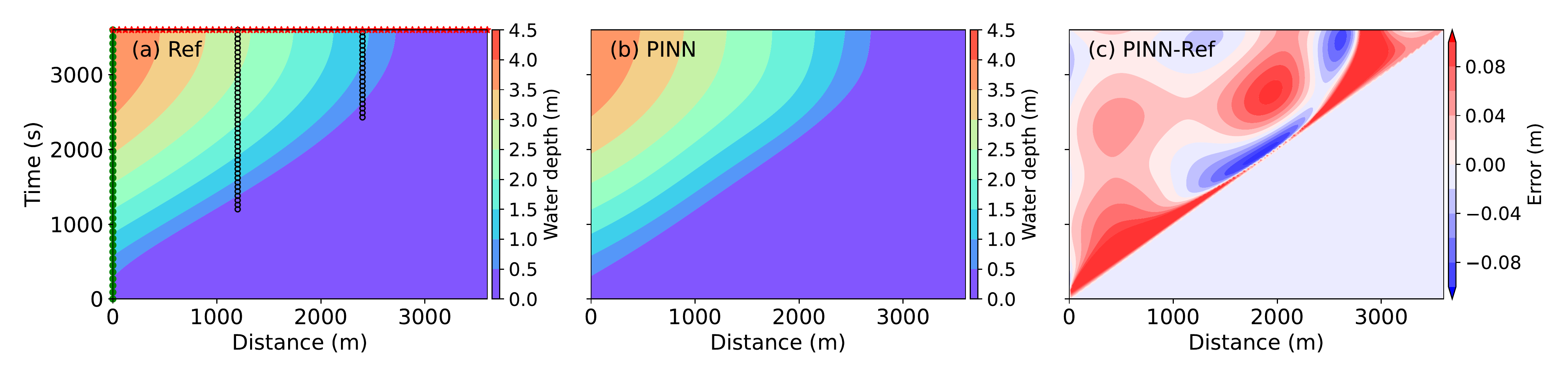}}
	\caption{The spatiotemporal evolution of the water depth in case 2 given by the RK4 solution (a), the PINN solution (b) and the error in the PINN solution (c). }
	\label{fig:case2_contour}
\end{figure}

\begin{figure}[htbp]
	\centering
	\includegraphics[width=\textwidth,keepaspectratio=true]{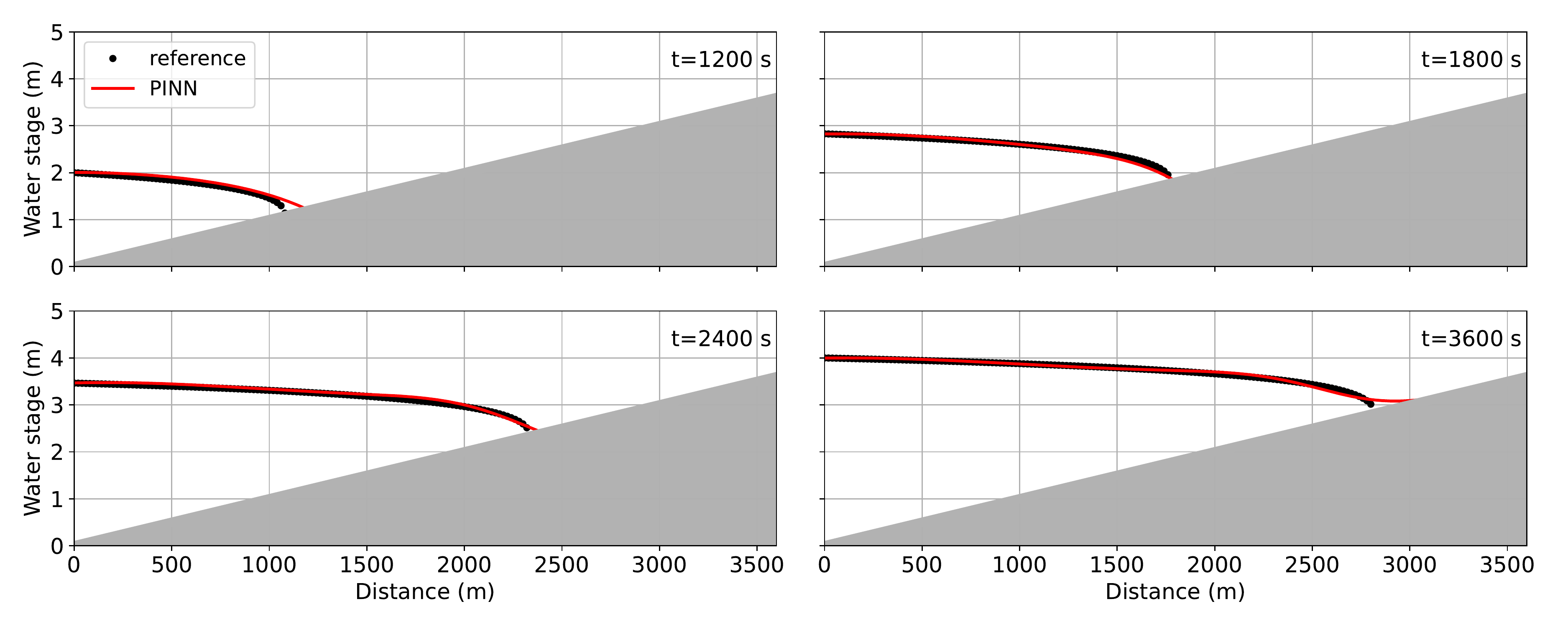}
	\caption{The water stage of the RK4 solution and the PINN solutions at four time instants in case 2.}
	\label{fig:case2_along_channel}
\end{figure}

The comparisons between the RK4 solution and the PINN solution over the spatiotemporal domain and along the travel distance are provided in Figure \ref{fig:case2_contour} and \ref{fig:case2_along_channel}, respectively. PINN is able to capture the flood propagation and the water surface profile well. The small bias only occurs near the end of the inundation front, which is most likely because there is no information enforced at the far end and the boundary of $h=0$ is specified at $x=ut$, a distance greater than the true travel distance. 
Overall, the resulting $\epsilon_h$ and RMSE imply reasonable accuracy of the PINN solution (Table \ref{table:case1234}).

In Case 1 and 2, we tested the PINN's capability of solving simplified SVE with a moving boundary, demonstrating the flood propagation over a floodplain can be well simulated by PINN. Also, provided with adequate observations and topology information, PINN can be used to compute the floodplain inundation at the subgrid scale. Even though our intention is not to replace the efficient inundation scheme of a large-scale river model that covers the entire domain \cite{luo2017modeling,yamazaki2011physically}, PINN can provide a more detailed inundation map at local regions of interest as the within-cell heterogeneity of water depth is resolved.

\subsubsection{Case 3 and 4}

\begin{figure}[htbp]
	\centering
	\makebox[\textwidth][c]{\includegraphics[width=\textwidth,keepaspectratio=true]{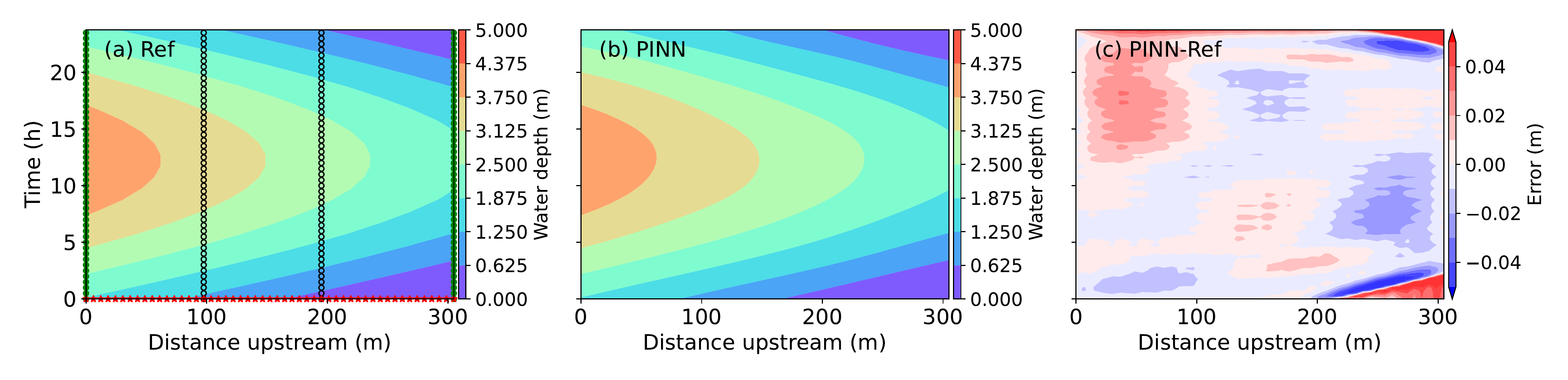}}
	\caption{The spatiotemporal evolution of the water depth in case 3 given by the HEC-RAS solution (a), the PINN solution (b) and the error in the PINN solution (c). }
	\label{fig:case3_contour}
\end{figure}

\begin{figure}[htbp]
	\centering
	\includegraphics[width=\textwidth,keepaspectratio=true]{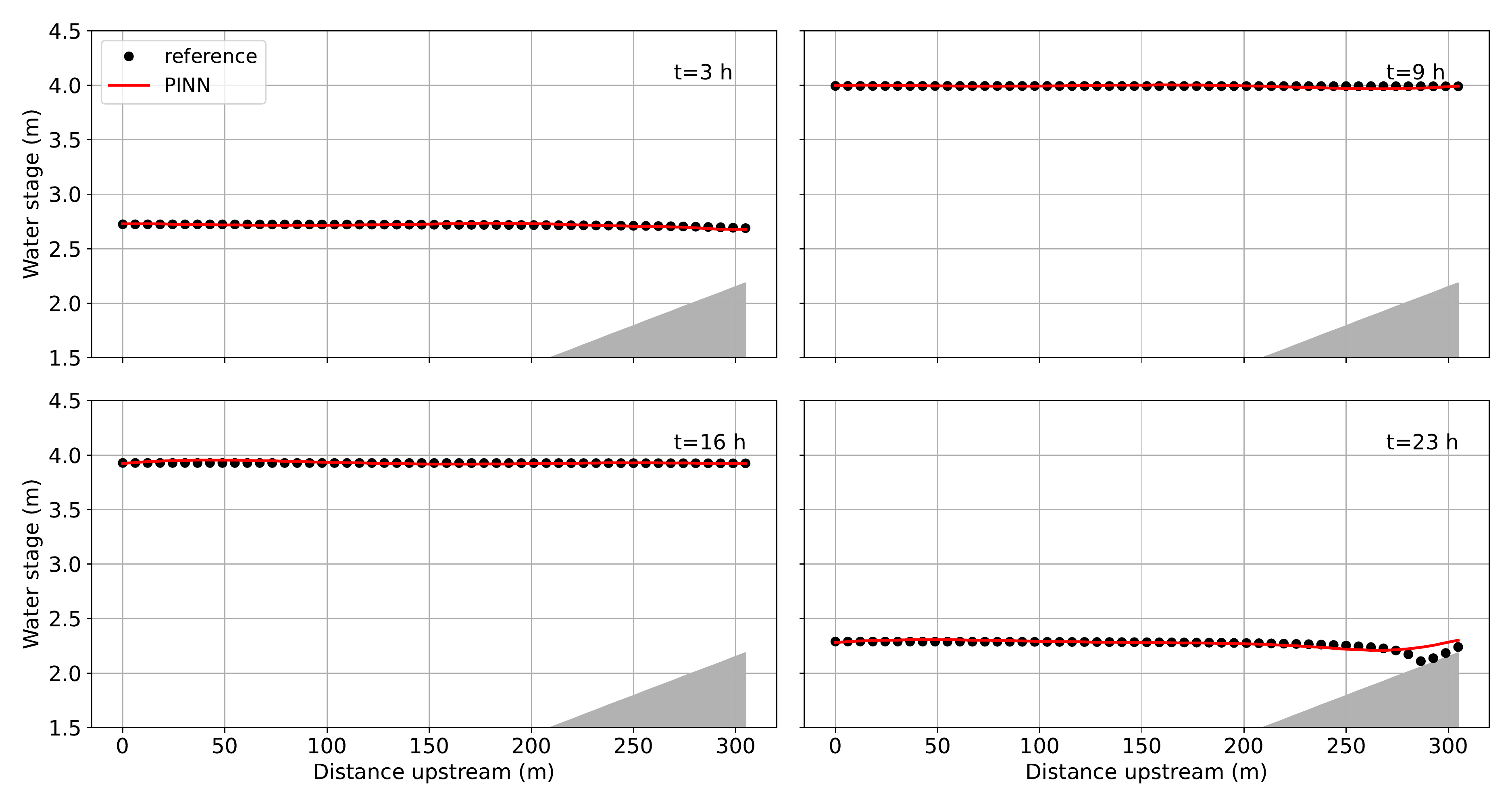}
	\caption{The along-channel water stage of the HEC-RAS solution and the PINN solutions at four time instants.}
	\label{fig:case3_along_channel}
\end{figure}

The PINN solutions are evaluated against the HEC-RAS reference solutions in terms of the water depth ($h$) over the spatiotemporal domain (Figure \ref{fig:case3_contour}) and their along-channel profiles at four different time instants (Figure \ref{fig:case3_along_channel}). The PINN solution agreed well with the reference solution, with small simulation bias near the upstream end of the along-channel profile.  
The corresponding $\epsilon_h$ and RMSE also imply good performance (Table \ref{table:case1234}). 

\begin{figure}[htbp]
	\centering
	\includegraphics[width=\textwidth,keepaspectratio=true]{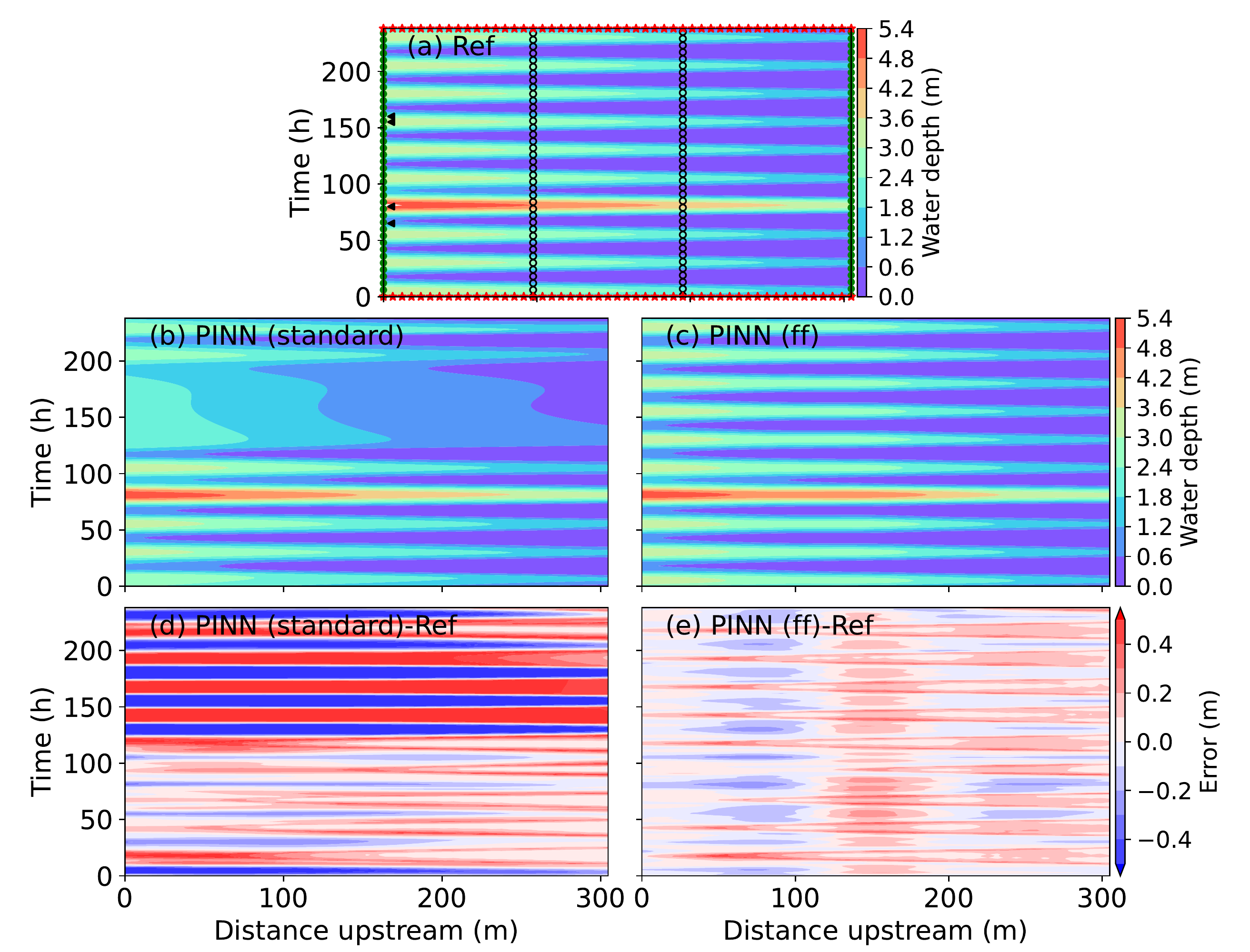}
	\caption{The spatiotemporal evolution of the water depth in case 4 given by the reference HEC-RAS solution (a), the standard PINN solution without Fourier feature embeddings (b) and the corresponding error (d), and the PINN solution with Fourier feature embeddings (c) and the corresponding error (e). }
	\label{fig:case4_contour}
\end{figure}

\begin{figure}[htbp]
	\centering
	\includegraphics[width=\textwidth,keepaspectratio=true]{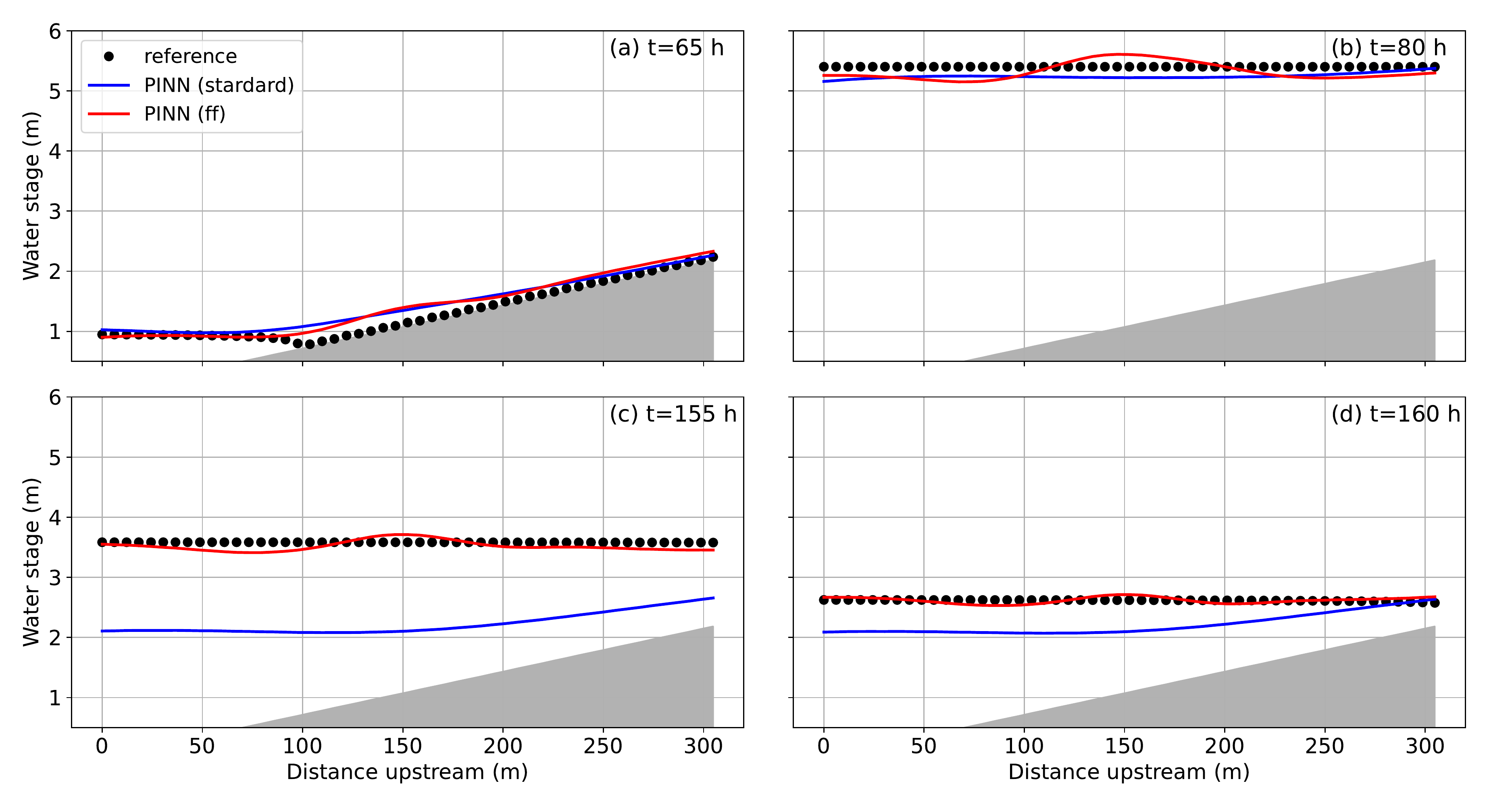}
	\caption{The along-channel water stage of the HEC-RAS solution and the PINN solution with and without Fourier feature embeddings at four time instants in case 4. The time instants are marked by the left triangles in Figure \ref{fig:case4_contour}(a). }
	\label{fig:case4_along_channel}
\end{figure}

Figures \ref{fig:case4_contour} and \ref{fig:case4_along_channel} compare the simulated spatiotemporal evolution of $h$ and the along-channel profiles between the HEC-RAS reference model and the two PINN models. The PINN solution that encodes the Fourier feature is able to capture the periodic variation induced by the downstream tide. The tidal phase and the amplitude of the tide and surge are reproduced (Figure \ref{fig:case4_contour}(c)). The rising and ebbing of the water surface are also well simulated. 
Figure \ref{fig:case4_along_channel} shows two time instants during the propagation of tide and storm surge and two time instants during the tidal ebbing.
The larger bias only occurs during low tides when water flows relatively fast towards the downstream end and between the two observational sites (Figure \ref{fig:case4_along_channel}). The biases at $x=100$ m and $x=200$ m are likely due to the weighting of the loss function in Equation \ref{eq:total_loss} that balances the observation loss ($J_{obs}$) with the other loss terms (i.e., $J_f$, $J_{BC}a$ and $J_S$).
In contrast, the PINN solution using the standard architecture can only predict the storm surge and a few tidal cycles and results in poor performance when the tidal variations are not captured (Figure \ref{fig:case4_along_channel}b and c).  
The solution with Fourier feature embeddings has a much smaller $\epsilon_h$ and RMSE than the standard PINN solution (Table \ref{table:case1234}). 
Overall, the accuracy of PINN prediction for case 4 is worse than the case 3, even with the enhancement by the feature encoding, because of higher oscillatory behaviours (see Figure 11a) created by tides and boundary conditions. This implies more physical constraints and observation data are required in the PINN training. The impact of assimilating more observations is discussed in Section \ref{s:case6}. 

\begin{table}[htbp]
	\caption{Information of the simulation errors in case 1$\sim$4. }
        \footnotesize
	\centering
	\begin{tabular}{ p{1.5 cm} p{3.2cm} p{2.2cm} p{2.0cm}  }
		\hline
		  Case No. & PINN scheme & $\epsilon_h$  & RMSE (m) \\
            \hline
            case 1                  & standard & 3.075$\times$10$^{-3}$ & 0.001          \\
            \hline 
		case 2                  & standard & 3.685$\times$10$^{-2}$ & 0.076          \\	
            \hline
            case 3                  & standard & 6.635$\times$10$^{-3}$ & 0.054 \\
		\hline
		\multirow{2}{*}{case 4} & standard & 3.093$\times$10$^{-3}$ & 1.771 \\
		                          & ff       & 7.176$\times$10$^{-2}$ & 0.411  \\	
		\hline
	\end{tabular}
	\label{table:case1234}
\end{table}

\subsubsection{Case 5 and 6}

\begin{figure}[htbp]
	\centering
	\includegraphics[width=\textwidth,keepaspectratio=true]{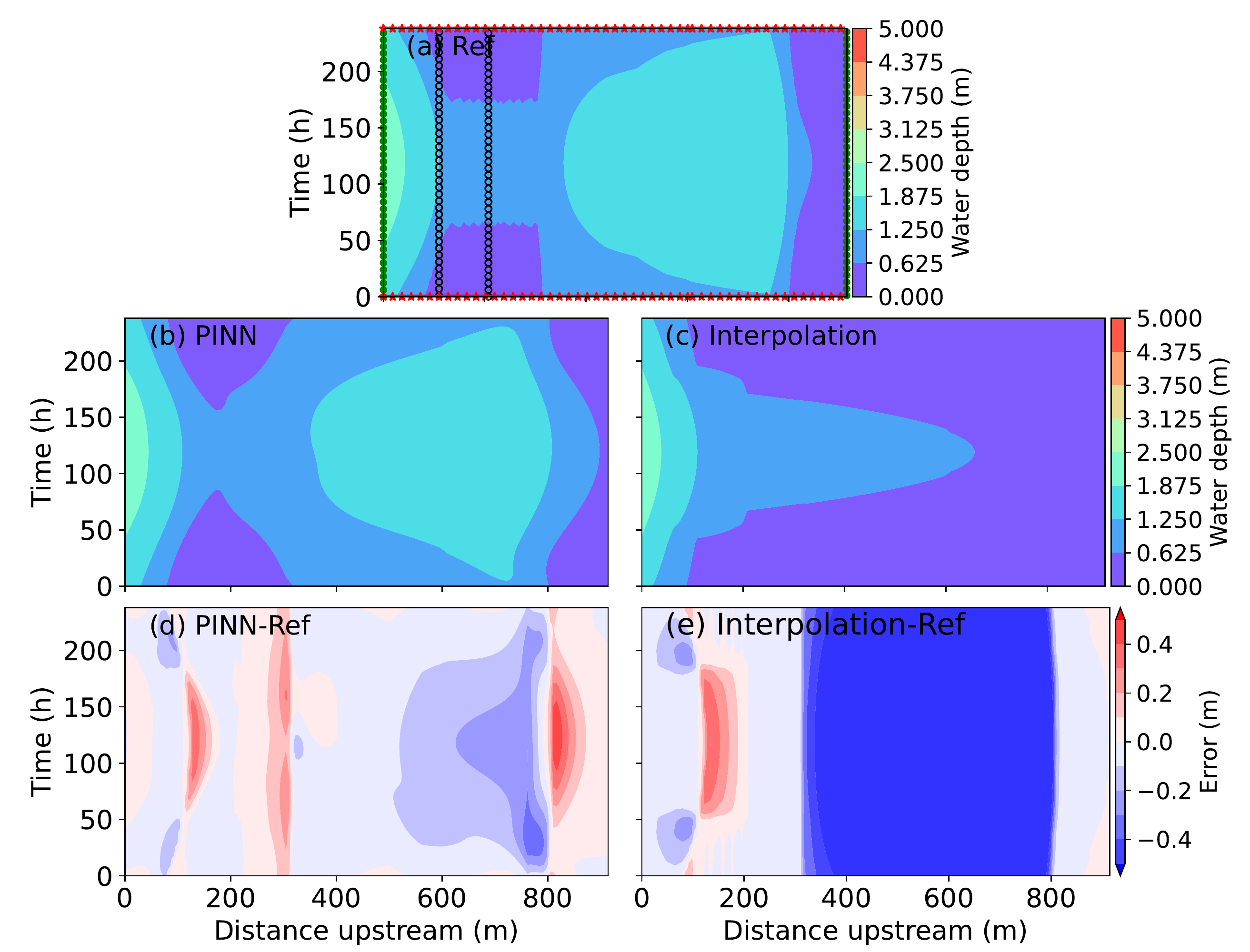}
	\caption{The spatiotemporal evolution of the water depth in case 5 given by the reference HEC-RAS solution (a), the PINN solution (b) and the corresponding error (d), the linear-interpolation solution (c) and the corresponding error (e). }
	\label{fig:case5_contour}
\end{figure}

\begin{figure}[htbp]
	\centering
	\includegraphics[width=\textwidth,keepaspectratio=true]{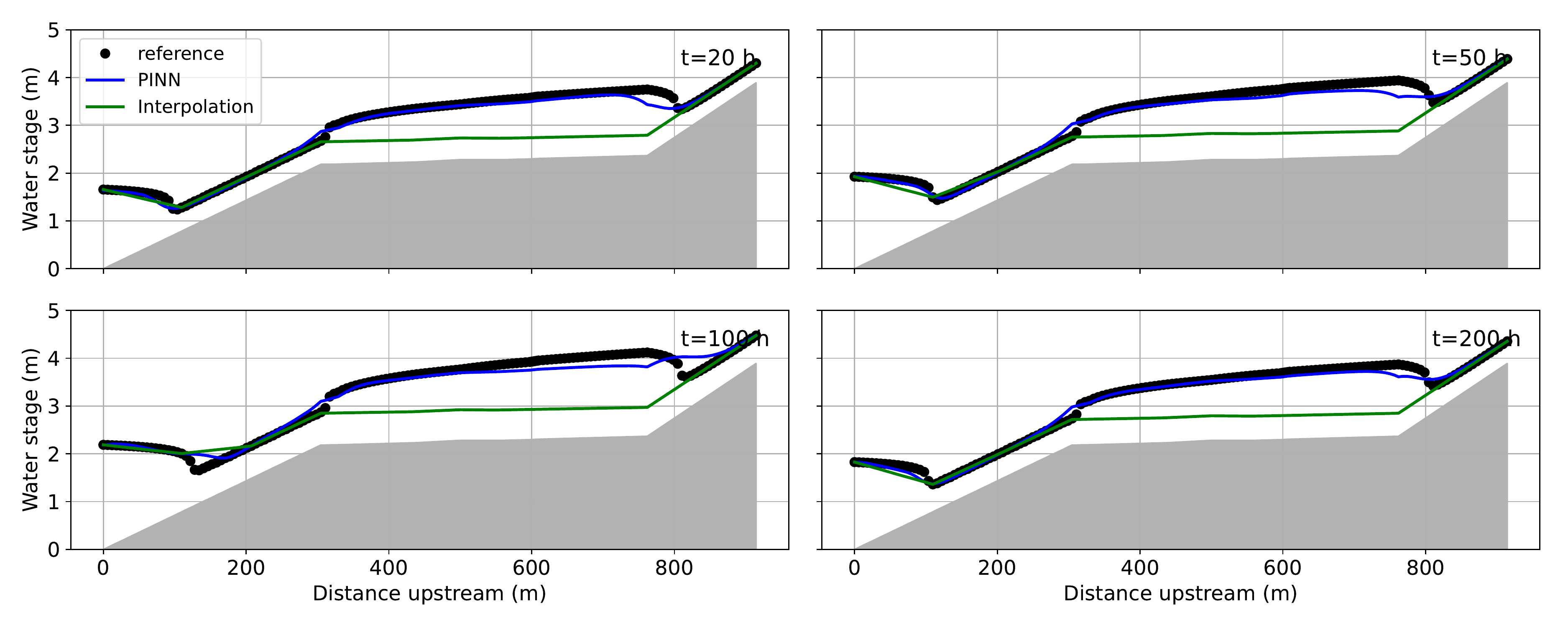}
	\caption{The along-channel water stage of the HEC-RAS solution, the PINN solution and the linear-interpolation solution at four instants of time in case 5.}
	\label{fig:case5_along_channel}
\end{figure}

The PINN predicted $h$ is compared with the HEC-RAS and linear-interpolation solutions over the full spatiotemporal space (Figure \ref{fig:case5_contour}) and along the channel at a few time instants (Figure \ref{fig:case5_along_channel}). 
The linear interpolation yields a poor downscaled solution, with the interpolated $h$ only agreeing with the reference over the downstream section. The accuracy of the linear interpolation reduces significantly in the other two sections where there are no in-situ data. Neither the spatiotemporal patterns nor the along-channel profiles are reproduced in these regions. The PINN solution, on the other hand, has a more resolved flow dynamics over the full subgrid sections. Particularly at the middle section where the subcritical flow dominates and the water depth is greater, PINN provides a more accurate solution (Figure \ref{fig:case5_along_channel}).  The errors of the PINN solution only arises where the flow regime changes.
In general, the errors (i.e., $\epsilon_h$ and RMSE) of the downscaled solution in PINN are less than 1/5 of those in the linear-interpolation solution. 

\begin{figure}[htbp]
	\centering
	\makebox[\textwidth][c]{\includegraphics[width=\textwidth,keepaspectratio=true]{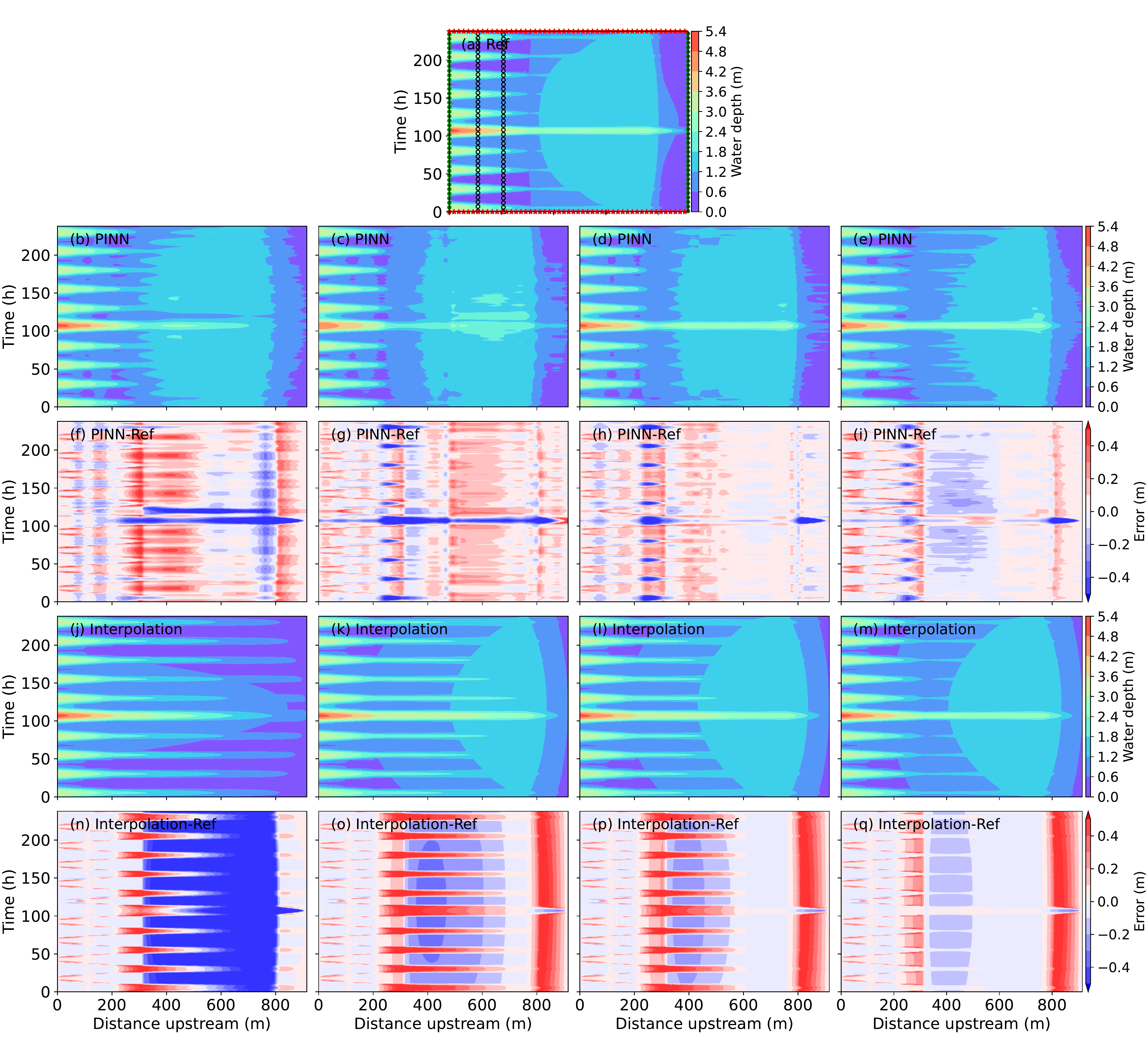}}
	\caption{The spatiotemporal evolution of the water depth in case 6. The HEC-RAS reference solution (a) is compared against the PINN solution (b)$\sim$(e) and the linear-interpolation solution (j)$\sim$(m). The errors in the PINN and linear-interpolation solutions are shown in (f)$\sim$(i) and (n)$\sim$(q), respectively.  The four columns correspond to the solutions with 2, 3, 4 and 5 observations. }
	\label{fig:case6_contour}
\end{figure}

\begin{figure}[htbp]
	\centering
	\includegraphics[width=\textwidth,keepaspectratio=true]{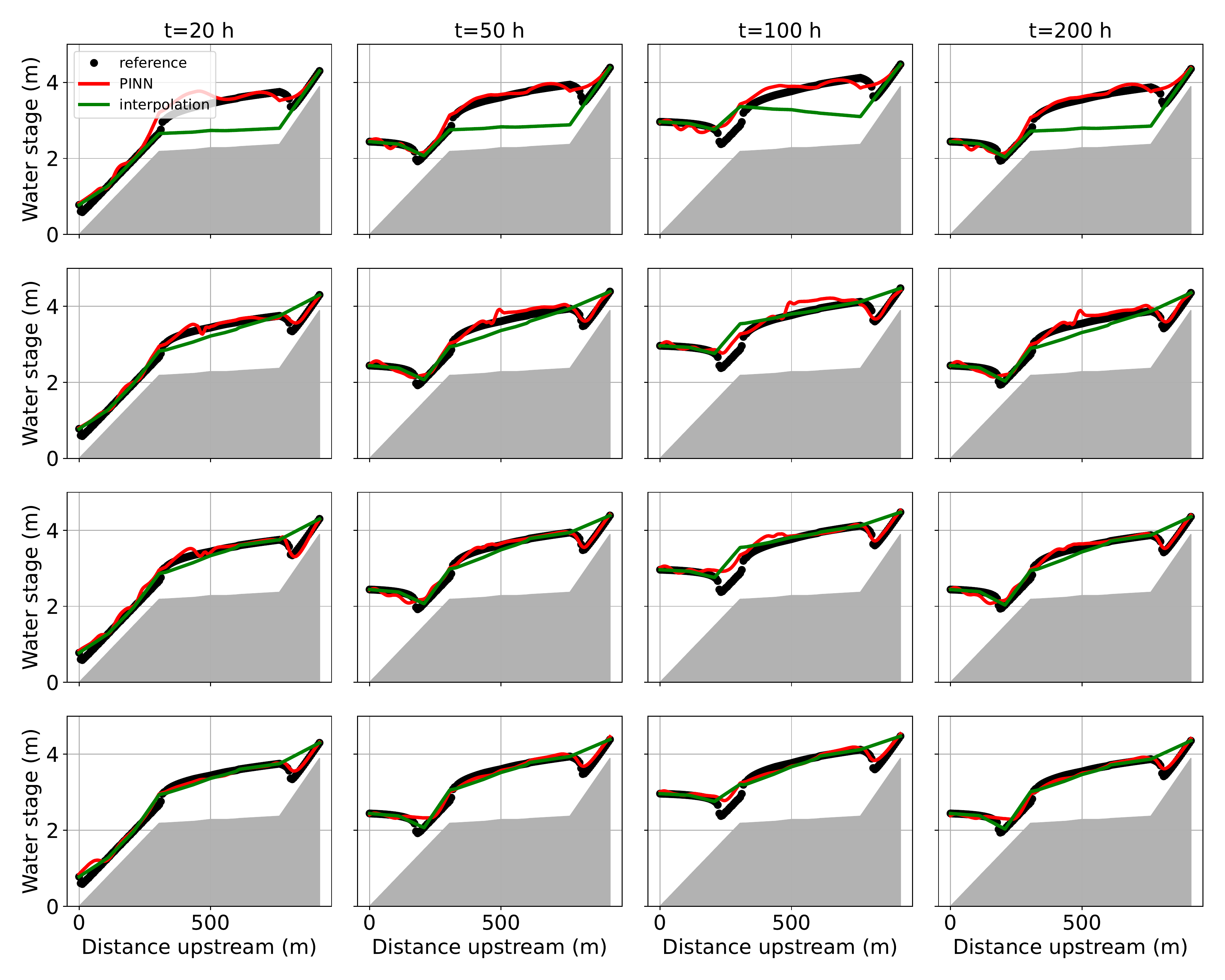}
	\caption{The along-channel water depth of the HEC-RAS solution, the PINN solutions and the linear-interpolation solutions at four time instants in case 6. The four panels correspond to the case of 2, 3, 4 and 5 observations in Table \ref{table:case56}, respectively. }
	\label{fig:case6_along_channel}
\end{figure}

The PINN solutions are compared with the HEC-RAS reference and the linear-interpolation solutions over the spatiotemporal space (Figure \ref{fig:case6_contour}) and along the channel (Figure \ref{fig:case6_along_channel}). While both downscaling methods show improved performance with the increase in in-situ data, the PINN solutions are more accurate and are less sensitive to the observations. 
When there are only gauges at the downstream section (x=0$\sim$305 m), PINN is able to capture the tidal phase and amplitude reasonably well (Figure \ref{fig:case6_contour}(b)). The periodic tide is characterized in the Fourier space, although there are small oscillations in the PINN solution. These oscillations are caused by the high-frequency component ($\textit{s}=10$) defined in the Fourier feature embedding.
The propagation of the storm surge and the gentle variation induced by the upstream discharge, however, are underestimated. In contrast, the linear interpolation overestimates the tidal influence in the two upstream sections (x=305$\sim$914 m), where the flow regimes cannot be resolved due to the lack of data (Figure \ref{fig:case6_contour}(c)). As a result, the linearly interpolated water depth is much smaller than the reference and PINN solutions (Figure \ref{fig:case6_along_channel}).
The error metrics also indicate that the PINN downscaling outperforms the linear interpolation (Table \ref{table:case56}): the errors are reduced to one third.   
When five in-situ stations are available (Figure \ref{fig:case6_contour}(h) and (i)), PINN still has more accurate solutions but the difference between the two methods is reduced. 
PINN shows slight improvements in the upstream stretch of the tides. The accuracy of the linear interpolation increases significantly over the two upstream sections. The water depth is better estimated except that near the upstream boundary which requires more data.   

The comparison of the two downscaling methods re-emphasizes the limitation of using the linear interpolation to downscale channel flow in variable flow regime conditions. This method has a large dependence on the amount of gauged data. Otherwise, the subgrid features cannot be resolved, resulting in poor downscaled solutions. 
In contrast, the PINN downscaling can achieve reasonable accuracy even when the observations are limited.

\begin{table}[htbp]
	\caption{Information of simulation errors in case 5 and 6. }
        \footnotesize
	\centering
	\begin{tabular}{ p{0.8cm} | p{1.0cm} p{3.0cm} p{3.0cm} p{2.2cm} p{1.2cm}  }
		\hline
		Case No. & Number of obs. & Location ($x$) & Downscaling method & $\epsilon_h$  & RMSE (m) \\
            \hline
\multirow{2}{*}{case 5} & \multirow{2}{*}{2} &  \multirow{2}{*}{100, 200}    & PINN                       & 9.876$\times$10$^{-2}$ & 0.387 \\
                        &                    &                               & Linear interpolation       & 5.181$\times$10$^{-1}$ & 2.033 \\	
            \hline
\multirow{8}{*}{case 6} & \multirow{2}{*}{2} &  \multirow{2}{*}{100, 200}    & PINN                       & 1.396$\times$10$^{-1}$ & 0.674 \\
		              &                    &                               & Linear interpolation       & 3.586$\times$10$^{-1}$ & 1.782  \\\cline{2-6}
		              & \multirow{2}{*}{3} & \multirow{2}{*}{100, 200, 764}  & PINN                       & 1.121$\times$10$^{-1}$ & 0.541 \\
		              &                    &                                 & Linear interpolation       & 1.410$\times$10$^{-1}$ & 0.873  \\\cline{2-6}
		              & \multirow{2}{*}{4} & \multirow{2}{*}{100, 200, 614, 764}  & PINN                       & 8.560$\times$10$^{-2}$ & 0.413 \\
		              &                    &                                      & Linear interpolation       & 1.288$\times$10$^{-1}$ & 0.763  \\\cline{2-6}
		              &\multirow{2}{*}{5} &  \multirow{2}{*}{100, 200, 314, 614, 764} & PINN                       & 8.142$\times$10$^{-2}$ & 0.393 \\
		              &                   &                                           & Linear interpolation       & 1.128$\times$10$^{-1}$ & 0.557  \\
		\hline
	\end{tabular}
	\label{table:case56}
\end{table}

\subsubsection{Computational effort}

The simulation time of the numerical and PINN solutions in Case 1$\sim$6 and the linear-interpolation downscaling in Case 5 and 6 are summarized in Table \ref{table:computational_time}. The simulations and linear interpolations are performed on an Intel Xeon CPU. 
The size of the problem and the NN structure regulate the PINN's computational time. It increases with the number of collocation points ($N_f$) and the training data ($N_{data}$), the NN size (e.g., number of hidden layers and neurons per layer) and the complexiblity in NN architecture. For example, the training time of the cases using the Fourier feature embeddings is over two times larger than that of those using the standard NN. 

PINN, despite having a higher computational cost than the numerical methods, shows promising performance when the training dataset has a relatively small size and the problem size is comparative to realistic conditions.  
More importantly, to obtain the subgrid solution of multiple grid cells, a local numerical model has to be configured separately for each corresponding cell. Considering such efforts, PINN can be more efficient and flexible than a local numerical model. 
PINN also has multiple advantages over the surrogate-type ML models in efficiency. Although the surrogate models may be faster after sufficiently training, the training of a surrogate model could require extensive efforts including a larger training dataset. The training data usually come from setting up and pre-running local numerical models for each cell of interest. In contrast, the PINN solutions can be obtained directly from the training process. 
Case 5 and 6 are designed at the size similar to realistic problems: the study domain is 900 m, i.e., the typical resolution in large-scale river models ($\sim$O(1km)); the domain is 
discretized to 6 m, i.e., the high resolution of SWOT (O(10m)); the 10-day simulation period is also determined based on the revisit time of satellite. 
Our results imply that the downscaled solution of the interested cell can be obtained in 10$\sim$50 min. This time is longer than a single numerical run. But PINN with its meshless feature reduces the effort of mesh generation in the numerical setup. Normally, the generated mesh in models, such as HEC-RAS, has to be adjusted manually multiple times prior to a successful run, which requires significant efforts when the downscaled cells are multiple. 
Moreover, the computational cost of PINN can be reduced using parallel computing and GPUs. 
While it is straightforward to deploy PINN to downscale multiple cells simultaneously, more sophisticate parallelism are now available \cite{meng2020ppinn,shukla2021parallel} to accelerate the process.

\begin{table}[htbp]
	\caption{Information of the computational time.}
        \footnotesize
	\centering
	\begin{tabular}{ p{0.8cm} | p{1.0cm} p{1.0cm} p{1.0cm} p{1.8cm} p{1.2cm} p{2.2cm} }
		\hline
		Case No. & scheme & $N_f$ & $N_{data}$ & Reference (s)  & PINN (s) & Interpolation (s) \\
            \hline
            case 1 & standard & 7380 & 605 & 0.012 & 440 & \\
            \hline
            case 2 & standard & 10980 & 605 & 0.184 & 620 & \\
            \hline
            case 3 & standard & 4896 & 339 & 21 & 158 & \\
            \hline
            \multirow{2}{*}{case 4} & standard & 12189 & 822 & 36 & 338 & \\
		              & ff & 12189 & 822 & 36 & 724 & \\
            \hline
            case 5 & standard & 36567 & 1262 & 38 & 615 & 0.016 \\
            \hline
            \multirow{4}{*}{case 6} & ff & 36567 & 1262 & 42 & 1529 & 0.018 \\
		            & ff  & 36567 & 1501 & 42 & 1546 & 0.018 \\
		        & ff  & 36567 & 1740 & 42 & 2438 & 0.018 \\
		          & ff  & 36567 & 1979 & 42 & 3147 & 0.018 \\
		\hline
	\end{tabular}
	\label{table:computational_time}
\end{table}

\subsection{Discussion of the test results}

The results in case 6 indicate that the availability of observations for assimilation plays an important role in determining the accuracy of the PINN solutions. 
With limited collocation points, the performance of PINN is degraded with increased variations of the state variables in space and time. For example, the temporal variations of water depth in case 4 are greater than those in case 3, resulting in a larger bias. Similarly, the spatially varied flow regime in case 6 causes the PINN to give less satisfied predictions than the uniform flow regime in case 4. This issue is related to a common challenge in training PINN. The solutions of the cells that are far from the domain boundaries and have drifted from the initial state can be easily trapped to a trivial local minimum during the PINN inference \cite{wong2021learning} because the constraints of the prescribed boundary and initial conditions on the solutions of those cells damp with time. In practical problems, additional constraints can be enforced by assimilating more observational data wherever available. Adding a time series of observations at a location between the boundaries of the computational domain is equivalent to subdividing the domain into subdomains in 1-D problems. 
The variations of each state variable become less significant in each subdomain, facilitating the learning process. 
However, even with spatially-varied flow regimes, PINN is less dependent on observations than the linear interpolation in downscaling 1-D channel flow. In practical applications, in-situ observations are likely limited to main gauging stations that are concentrated near the coastal zone (e.g., the downstream section in case 6). In such a case, PINN is still able to provide reasonable downscaled solutions due to its capability of encoding the governing physics to resolve the topology and assimilating the remote sensing snapshots. The resolved variation of the water depth is of critical importance for delineating more accurate flood inundation maps.

Our proposed Fourier feature embeddings are sufficient to encode the physical characteristics of tide and address the periodicity in BC. However, the PINN solution in case 6 suggests that the performance is degraded when the flow varies dramatically along the channel. 
Even though the periodic tidal variations are formulated into the Fourier space, the along-channel dynamics is problem-dependent, as the flow regime is determined by various factors, including the channel topology, the upstream and downstream BCs and their interactions. In case 6, while the flow is determined by the high-frequency tide in the downstream section, the variation of flow is much smaller in the upstream and middle sections, where the tidal influence is limited and the low-frequency discharge dominates.
Because Fourier transformation may only be performed on partial training data, it is challenging to encode the space variable directly in the NN architecture. Moreover, Fourier feature embeddings modulate the variability in data to match the high-frequency patterns. As such, the nonlinear interactions between high-frequency and low-frequency processes are not well reproduced, especially when there are limited observations. In case 6 for example, the PINN solutions have larger biases near the tips of the tidal propagation where the discharge-tide interaction occurs (Figure \ref{fig:case6_contour}). Such interaction is expected to be stronger at the upstream reaches of real tidal rivers, and more complex than what we represent in the case since other mechanisms also occur in the context of compound flooding \cite{dykstra2020propagation}. This poses further challenges in the realistic application of PINN.  Fortunately, the accuracy of PINN depends on the training data. That means, if the proposed PINN solves SVE with sufficient data assimilated, the flow dynamics and discharge-tide interaction would be reasonably emulated. Our results show the accuracy improved with the increase in the amount of assimilated observations. The PINN-based data assimilation provides a new path towards tackling the challenges of compound flood modeling in large-scale earth system models. 

The most valuable insight developed in this work is a relatively efficient way of downscaling a coarse-scale model at the subgrid domain using the physics-based ML/AI techniques. The PINN-based downscaling is proved quite handy for assimilating both in-situ and remote sensing data. The latter, in particular, provides a key addition to the numerical modeling community for better resolving the physics without generating subgrid meshes and modifying the model code, thus preserving the computational efficiency. 
The downscaling method is mesh independent and can always be used to solve a subgrid solution for the interested sub-regions of a large domain, regardless of the original mesh resolution. With the increase in data abundance, the method enables the investigation of subgrid physics even in a high-resolution ($\sim$O(1m)) model, alleviating limitations in numerical modeling.
The downscaling discussed herein provides us a promising path to integrate advanced remote sensing to dynamic numerical models.  

\subsection{Limitations and future applications of PINN}

In the future, we aim at applying the PINN downscaling to a broader domain with more realistic flow conditions. Our test cases are relatively simple. First, we only apply a constant friction coefficient ($n$). However, $n$ may be considered as a trainable variable in NN because the variable $n$ is often used to account for enhanced friction effects from the nonlinear interaction among tides, storm surge and river discharge \cite{godin1991some,godin1999propagation,kukulka2003impacts}. Second, the effects of river channel geometry including the channel meandering and widening towards the coast have not been considered either. The latter, in particular, has been found to increase the channel capacity and flow gradient \cite{fuller2013methods}. Additionally, general issues, such as observation uncertainty, inconsistent data time series and lack of observations, are common in realistic conditions.
While intermittent data are applicable in training without any modifications, a more robust noise-resistance algorithm will be needed to handle data uncertainties or errors . In this study, the noise level is set by USGS in-situ measurement errors. However, the remote sensing data errors are subject to large variability. While the local-scale remote sensing measurements can achieve high accuracy \cite{bhattacharjee2021accuracy}, the large-scale satellite images may suffer larger errors \cite{biancamaria2016swot}. Finally, it is necessary to improve the accuracy of PINN in resolving the nonlinear interaction at the river-ocean interface. 
Overall, the application of PINN in realistic conditions is challenging and has not yet been explored extensively. Even though we have tested and developed algorithms best fitting our purpose, in the future we will also explore state-of-the-art physics-based machine learning techniques and tools that are emerging in the ML community, such as the general PINN library DeepXDE \cite{lu2021deepxde}, the Fourier Neural Operator that models turbulent flows \cite{li2020fourier} and algorithms that could potentially improve the PINN robustness to noise \cite{bajaj2021robust}.

This study demonstrates the PINN's capability of solving the 1-D SVE by assimilating data. However, the 1-D approach neglects the vertical stratification and baroclinic forcing during a storm tide propagation \cite{orton2012detailed} and limits the application in 2-D flood mapping. As the SWOT satellite imagery provides both 1-D vector and 2-D raster products for water surface elevations, we expect to extend the PINN-based data assimilation approach to a 2-D domain. Even though the 2-D application will bring in the velocity in a different direction and stimulate more interactions, 
the success of PINN in solving Navier-Stokes equations \cite{rao2020physics,jin2021nsfnets} provides the theoretical basis for tacking 2-D river dynamics in both depth-averaged and longitudinal-averaged form. 

In addition to being integrated in a numerical model, PINN can also be used to create a digital twin targeting on regions with complex flow regimes, where large-scale models at their current state are unable to resolve. 
For example, at the river-ocean interface, 1-D hydraulic models (e.g. HEC-RAS) are usually employed. The hydraulic models take freshwater discharge from hydrologic models at the upstream boundary and water stage information from ocean models at the downstream boundary \cite{torres2015characterizing}. These models are more computational efficient than 2-D/3-D hydrodynamic models and resolve the flow dynamics and topography better than hydrologic models. 
Such hydraulic models may be replaced by a digital twin based on PINN, to exchange the information from both sides at the river-ocean interface. Such a digital twin can be developed following a similar procedure as shown in case 6: (i) the BC data are obtained from the linked river and ocean models at the domain boundaries; (ii) the snapshots are provided by satellite images and/or other remote sensing data; (iii) in-site measurements are applied as additional constraints.

\section{Conclusion}
\label{s:conclusion}

This research makes the first attempt to apply the state-of-the-art PINN to downscale a large-scale river model to a subgrid solution. The conservation physics of the governing equations regularizes the training dataset to a manageable size, which alleviates the burden in data-intensive neural network models. The proposed PINN framework can (i) solve the full dynamic 1-D SVE equation, (ii) encode the periodic variation in tidal boundaries using Fourier feature embeddings, (iii) assimilate various data types and (iv) be integrated to a large-scale river model for downscaled solutions. This work tests the applicability of PINN in flood modeling for delineating more accurate inundation maps, and creates a promising path for applying the method to real-world problems with the emergence of high-quality satellite images in surface water storage and fluxes and more comprehensive in-situ gauge networks.


\section*{Acknowledgments}
The work presented in this manuscript is supported by the Earth System Model Development program areas of the U.S. Department of Energy, Office of Science, Office of Biological and Environmental Research as part of the multi-program, collaborative Integrated Coastal Modeling (ICoM) project. PNNL is operated for DOE by Battelle Memorial Institute, United States under contract DE-AC05-76RL01830.

\section{Code and data availability}
The data and code to reproduce the results and figures are publicly available at the Zenodo repository with identifier 10.5281/zenodo.7118168.


\bibliographystyle{unsrt}  
\bibliography{references}

\end{document}